\documentclass[aps,twocolumn,floatfix]{revtex4}

\usepackage{amsmath,amsthm,amsfonts,amssymb,times,bbm,graphicx}

\newtheorem{theorem}{Theorem}

\newtheorem{lemma}[theorem]{Lemma}
\newtheorem{corollary}[theorem]{Corollary}
\newtheorem{definition}[theorem]{Definition}

\def\Symp#1,#2,#3,#4.{\left[\left(\begin{array}{c}#1\\#2\end{array}\right),\left(\begin{array}{c}#3\\#4\end{array}\right)\right]}
\def\Vec#1,#2.{\left(\!\begin{array}{c}#1\\#2\end{array}\!\right)}
\def\tuple#1.{\langle#1\rangle}
\def\ket#1.{|#1\rangle}
\def\bra#1.{\langle#1|}
\def\braket#1,#2.{\langle#1|#2\rangle}
\def\Gauss#1,#2.{{#1\brack#2}_d}

\def\ZZ{\mathbbm{Z}}
\def\RR{\mathbbm{R}}
\def\CC{\mathbbm{C}}
\def\FF{\mathbbm{F}}
\def\NN{\mathbbm{N}}

\def\H{\mathcal{H}}
\def\Fou{\mathcal{F}}
\def\Id{\mathbbm{1}}
\def\spacedot{\,\cdot\,}

\newcommand{\operp}{\ensuremath{\text{\textcircled{$\perp$}}}}

\DeclareMathOperator{\tr}{tr}
\DeclareMathOperator{\Tr}{Tr}

\DeclareMathOperator{\supp}{supp}
\DeclareMathOperator{\const}{const}

\DeclareMathOperator{\ord}{ord}
\DeclareMathOperator{\minarg}{minarg}
\DeclareMathOperator{\Stabs}{Stabs}
\DeclareMathOperator{\Iso}{Iso}

\begin{document}

\title{Hudson's Theorem for finite-dimensional quantum systems}

\author{D. Gross}

\affiliation{
	Institute for Mathematical Sciences, Imperial College London, London
	SW7 2BW, UK
} 
\affiliation{
	QOLS, Blackett Laboratory, Imperial College London, London SW7 2BW,
	UK
}

\email{davidg@qipc.org}

\date{\today}

\begin{abstract} 
We show that, on a Hilbert space of odd dimension, the only pure
states to possess a non-negative Wigner function are stabilizer
states. The Clifford group is identified as the set of unitary
operations which preserve positivity.  The result can be seen as a
discrete version of Hudson's Theorem.  Hudson established that for
continuous variable systems, the Wigner function of a pure state has
no negative values if and only if the state is Gaussian.  Turning to
mixed states, it might be surmised that only convex combinations of
stabilizer states give rise to non-negative Wigner distributions. We
refute this conjecture by means of a counter-example. Further, we give
an axiomatic characterization which completely fixes the definition of
the Wigner function and compare two approaches to stabilizer states
for Hilbert spaces of prime-power dimensions. In the course of the
discussion, we derive explicit formulas for the number of stabilizer
codes defined on such systems.
\end{abstract}


\maketitle

\section{Introduction}

\subsection{General Introduction}

The Wigner distribution establishes a correspondence between quantum
mechanical states and real pseudo-probability distributions on phase
space. 'Pseudo' refers to the fact that, while the Wigner function
resembles many of the properties of probability distributions, it can
take on negative values. This phenomenon has been linked to
non-classical features of such quantum states (see Ref. \cite{zyko}
for an exposition of literature on that problem). It is naturally of
interest to characterize those quantum states that are classical in
the sense of giving rise to non-negative phase space distributions.

For the case of pure states described by vectors in $\H=L^2(\RR)$, the
resolution of this problem was given by Hudson in Ref.
\cite{hudson}. Later, Soto and Claverie generalized Hudson's result to
states of multi-particle systems (Ref. \cite{soto}).

\begin{theorem}\label{thHudson}
	\emph{(Hudson, Soto, Claverie)}
	Let $\psi\in L^2(\RR^n)$ be a state vector. The Wigner function of
	$\psi$ is non-negative if and only if $\psi$ is a \emph{Gaussian
	state}.
	
	By definition, a vector is Gaussian if and only if it is of the form
	\begin{equation*}
		\psi(q)\propto e^{2\pi i (q \theta q+x q)},
	\end{equation*}
	where $q,x\in\RR^n$ and $\theta$ is a symmetric matrix with entries in
	$\CC$
	\footnote{
		Note that the boundedness of $\psi\in L^2(\RR^n)$ implies that
		$\theta$ has positive semi-definite imaginary part. 
	}.
\end{theorem}

It is our objective to prove that the situation for discrete quantum
systems is very similar, at least when the dimension of the Hilbert
space is odd. Before stating the result, we pause for a brief overview
of its main ingredients: discrete Wigner functions and stabilizer
states. 

The  Wigner function \cite{wigner} of a
pure state $\psi \in L^2(\RR)$ is computed as
\begin{equation}\label{wignerDef}
	W_\psi(p,q) = \pi^{-1} \int_{\xi\in \RR} e^{-2\pi i\xi p}\, 
	\bar\psi(q-\frac12 \xi)\psi(q+\frac12 \xi).
\end{equation}
Equivalently, $W_\psi$ is the (symplectic) Fourier transform of the
\emph{characteristic function} $\Xi_\psi$, which in turn is defined by
\begin{equation*}
	\Xi_\psi(p,q)=\tr(w(p,q)^\dagger \ket\psi.\bra\psi.).
\end{equation*}
Here, $w(p,q)=e^{i (p \hat X-q \hat P)}$ are the well-known
\emph{Weyl} or \emph{displacement operators}
\cite{wallsmilburn,folland}.  Partly triggered by the advent of
quantum information theory, considerable work has been undertaken to
explore Wigner functions for finite-dimensional quantum systems
\cite{oldWootters,vourdas,leonhardt,miquel,villegar,klimov,ruzzi,
chaturvedi,diploma,wootters}.
Two approaches might be identified in the literature on that subject.
The first one aims to cast the \emph{definition} of the Wigner
function into a form that can be interpreted for both continuous
variable and discrete systems \cite{vourdas,miquel,villegar,diploma}.
The second approach -- introduced by Gibbons, Hoffman, and Wootters in
Ref. \cite{wootters} -- focuses on the \emph{properties} of Eq.
(\ref{wignerDef}).  The authors imposed a set of axioms which a
candidate definition of a discrete Wigner function would have to
fulfill in order to resemble the well-known continuous counterpart.

We will argue that, for odd dimensions $d$, 
\begin{equation*}
	W_\psi(p,q) = d^{-1} \sum_{\xi\in \ZZ_d} e^{-\frac{2\pi}d i\xi p}\, 
	\bar\psi(q-2^{-1} \xi)\psi(q+2^{-1} \xi)
\end{equation*}
is the most sensible analogue of Eq.  (\ref{wignerDef}), judged in
terms of either of these approaches.  Here, $p,q$ are elements of
$\ZZ_d=\{0,\dots,d-1\}$ and $2^{-1}=(d+1)/2$ is the multiplicative
inverse of $2$ modulo $d$. Indeed, the definition given above is the
discrete symplectic Fourier transform of the discrete characteristic
function and will be shown to be the \emph{unique} choice to mimic
certain desirable properties of the continuous Wigner function.

Stabilizer states were originally defined by Gottesman in Ref.
\cite{gottesman} as the joint eigenvectors of certain sets of elements
of the qubit Pauli group.  Exceeding the case of qubits, stabilizer
states for higher-dimensional quantum systems have been treated in the
literature (see, e.g. Refs. \cite{knill,schlinge,grassl,erik}).  Such
states find manifold applications in quantum information theory,
ranging from quantum error correction \cite{nielsen} to Cluster state
quantum computation \cite{raussendorf}.  Although displaying complex
features such as multi-particle entanglement \cite{jensGraph},
stabilizer states allow for an efficient classical description. In
particular, a quantum computer that operates only with stabilizer
states can offer no principal advantage over classical methods of computing
\cite{nielsen}. The latter statement is sometimes called
\emph{Gottesman-Knill Theorem}.

Using that language, we intend to show:

\begin{theorem} \emph{(Discrete Hudson's Theorem)}\label{thMain}
	Let $d$ be odd and $\psi\in L^2(\ZZ_d^n)$ be a state vector.  The
	Wigner function of $\psi$ is non-negative if and only if $\psi$ is a
	\emph{stabilizer state}.

	Given that $\psi(q)\neq0$ for all $q$, a vector $\psi$ is a
	stabilizer state if and only if it is of the form
	\begin{equation*}
		\psi(q)\propto e^{\frac{2\pi}{d} i (q \theta q+x q)},
	\end{equation*}
	where $q,x\in\ZZ_d^n$ and $\theta$ is a symmetric matrix with entries in
	$\ZZ_d$.
\end{theorem}

Theorem \ref{thMain} should convey two central messages.  Firstly, if
the right definitions are employed, the continuous and the discrete
case behave very similarly (even though the methods of proof are
completely different). Secondly, it adds further evidence to what
might be called a piece of folk knowledge in the field of quantum
information theory: namely that stabilizer states are the natural
finite-dimensional analogue of Gaussian states.

The paper is organized as follows. We survey previous work on the
subject in Section \ref{scPrevious}.  Section \ref{scPhaseSpace} is
devoted to a superficial, yet self-contained introduction to 
Weyl operators, characteristic functions, Wigner distributions and
stabilizer states.  The main theorem is proven in Section
\ref{scMain}.  Sections \ref{scMixedStates} to \ref{scPrimePower}
address various related topics. The results of these last three
sections do not rely on each other. Concretely, we comment on the
relation between stabilizer states and Gaussian states in Section
\ref{scDiscreteGaussians}; we consider mixed states with positive
Wigner functions in Section \ref{scMixedStates} and use Section
\ref{scPrimePower} for a discussion of Hilbert spaces whose dimension
is the power of a prime.

Readers interested only in the structure of the proof, but not in
its full generality, are deferred to Ref.\ \cite{prime}, where a
particularly simple special case of the main result is laid out.

\subsection{Previous Results}
\label{scPrevious}

Recently, Galvao \emph{et. al.} took a first step into the direction
of classifying the quantum states with positive Wigner function 
\cite{galvao}. To explain the relationship of their
results to the present paper, we have to comment on 
an axiomatic approach to discrete Wigner functions and, further, on
stabilizer states in dimensions that are the power of a prime number.

In Ref. \cite{wootters}, Gibbons, Hoffmann, and Wootters listed a set
of requirements which should be met by any definition of a discrete
Wigner function $W$. Denoting the dimension of the Hilbert space by
$d$, their axioms amount to
\begin{enumerate}
	\item \emph{(Phase space)}
	$W$ is a linear mapping sending operators to functions on a
	$d\times d$ lattice, called the \emph{phase space}.

	\item
	\emph{(Translational covariance)}
	The Wigner function is covariant under the action of the Weyl
	operators (in the sense of Theorem \ref{thWignerCovariance}).

	\item \emph{(Marginal probabilities)}
	There exists a function $Q(\lambda)$ that assigns a pure quantum
	state to every line $\lambda$ in phase space. If $\psi$ is state
	vector, then the sum of its Wigner function along $\lambda$ must be
	equal to the overlap $|\braket Q(\lambda), \psi.|^2$.
\end{enumerate}
Let us call functions that fall into this class \emph{generalized
Wigner functions}. This term is justified, as the
characterization does not specify a unique solution: for a
$d$-dimensional Hilbert space, there exist $d^{d+1}$ distinct
generalized Wigner functions. Note also that the construction has been
described only for the case where $d=p^n$ is the power of a prime,
because only then the notion of a \emph{line} in phase space has a
well-defined meaning. 

We turn to the second remark, concerning stabilizer states.
Consider a composite system, built of $n$ $d$-level particles. We
are free to conceive it as a single $d^n$-dimensional object.  The two
points of view give rise to different definitions of stabilizer
states, the 'single-particle' one being starkly reduced as compared to
the multiple-particle one. In Section \ref{scPrimePower}, we show that
the set of single-particle stabilizer states is strictly contained in
the set of multi-particle ones. Indeed, the ratio of the respective
cardinalities of the two sets grows super-exponentially in $n$. 
As an example, the generalized Bell and GHZ states
\begin{equation*}
	d^{-n/2}\sum_i \ket i.\otimes\ket i.,
	\quad\quad 
	d^{-n/2}\sum_i \ket i.\otimes\ket i.\otimes \ket i.,
\end{equation*}
arguably the best-known multi-particle stabilizer states, do not belong to 
the respective single-particle sets.

The result of Ref. \cite{galvao} concerns quantum states in
prime-power dimensions that are non-negative with respect to
\emph{all} possible definitions of generalized Wigner functions. These
states are shown to be mixtures of single-particle stabilizer states,
as described above.  The authors aim to establish necessary
requirements for quantum computational speedup. Indeed, if the Wigner
function of a quantum computer is positive at all times, then it
operates only with stabilizer states and hence offers no advantage
over classical computers, by the Gottesman-Knill Theorem.

Thus for the case of non-qubit pure states, Theorem \ref{thMain}
implies the results of Ref. \cite{galvao} and goes further in two
essential ways.  Firstly, it suffices to look at a single definition
of the Wigner function, as opposed to $d^{n(d^n+1)}$ generalized ones.
Secondly, quantum computation and the Gottesman-Knill Theorem are
naturally set in the context of \emph{multiple} particles. Our
definition assigns positive Wigner functions to all multiple-particle
stabilizer states, while Ref. \cite{galvao} effectively relies on the
single-particle definition \footnote{
	Up to equivalence under Clifford operations.
}. On the other hand, our main theorem does
not address qubits or mixed states, which Galvao \emph{et.  al.} do.

\section{Phase Space Formalism}
\label{scPhaseSpace}

The term \emph{phase space formalism} encompasses the ideas and tools
in relation to the \emph{Weyl representation}, to be defined shortly.
We will give a concise introduction in this section. Many of the
results presented can be found in the literature, but some, e.g. the
Clifford covariance of the Wigner function in non-prime dimensions,
seem to be new.

\subsection{Weyl representation}

We start by considering a $d$-dimensional quantum system, $d$ odd.  In
its Hilbert space $\H$, we choose a basis $\{\ket0.,
\dots,\nolinebreak{\ket d-1.}\}$, labeled by elements of $\ZZ_d$.
Henceforth, $\ZZ_d$ will be referred to as the \emph{configuration
space} and abbreviated by $Q$.

The pivotal objects in the phase space formalism are the \emph{Weyl
operators} (also known as the \emph{generalized Pauli operators}), as
constructed below. Let $\chi(q)=e^{\frac{2\pi}d i q}$.  The relations
\begin{eqnarray}\label{shiftClock}
  \hat x(q)\ket x. = \ket x+q., \quad\quad
	\hat z(p)\ket x. = \chi(p x) \ket x.
\end{eqnarray}
define the \emph{shift} and \emph{boost} operators respectively.
The Weyl operators are given by
\begin{eqnarray}\label{weylOps}
	w(p,q)=\chi(-2^{-1}p q)\, \hat z(p)\hat x(q),
\end{eqnarray}
for $p,q,t \in Q$. The specific choice of phases will prove useful
later on \footnote{
	The choice of phase factors ensures that the symplectic inner
	product Eq. (\ref{symp}) appears in the composition law
	Eq. (\ref{heisenbergComp}) thus making the connection between the Weyl
	operators and symplectic geometry manifest. Other definitions
	in use, e.g. $w(p,q)=\hat z(p)\hat x(q)$ carry the same dependence
	in a less obvious manner.
	See also Refs. \cite{vourdas,folland}. 
}.
The set of Weyl operators is closed under multiplication, up to phase
factors.  Direct computation shows
that the composition law is given by
\begin{eqnarray}\label{heisenbergComp}
	&& w(p, q) w(p', q') \\
	&=&
	\chi(2^{-1} \Symp p, q, p', q'.)\,w(p + p', q + q'). \nonumber
\end{eqnarray}
The square brackets denote the standard \emph{symplectic 
inner product} on $\ZZ_d^2$:
\begin{eqnarray}\label{symp}
  \Symp p,q,p',q'.:=\Vec p, q.^T J \Vec p', q'.
\end{eqnarray}
where
\begin{equation} \label{jMatrix}
	J = 
		\left(
			\begin{array}{cc}
				0	&	1 \\
				-1 & 0
			\end{array}
		\right).
\end{equation}
We write $w(v)=w(v_p, v_q)$ for elements $v=(v_p,v_q) \in \ZZ_d^2$.
The space $V:=Q\times Q$ with inner product given by Eq. (\ref{symp})
will be called \emph{phase space} in the sequel, owing to its analogy
to the phase space known in classical mechanics.

The preceding constructing generalizes naturally to multiple particles. 
Indeed, the configuration space of an $n$-particle system is given by
$Q=\ZZ_d^n$. Multiplication between two elements $p,q\in Q$ is
understood as the usual inner product $pq = \sum_i p_i q_i$. The
Hilbert space is again spanned by $\{\ket q.\}_{q\in Q}$ and the Weyl
operators are defined to be the tensor products
\begin{eqnarray}\label{multipleWeyl}
	w(p,q) 
	&=& w(p_1, \dots, p_n, q_1, \dots, q_n) \\
	&=& w(p_1, q_1)\otimes \dots \otimes w(p_n, q_n).\nonumber
\end{eqnarray}
Equations (\ref{heisenbergComp}), (\ref{symp}) remain valid in the
multiple-particle setting, if we substitute the matrix $J$ by its
multi-dimensional version
\begin{equation*}
	J=
	\left(
		\begin{array}{cc}
			0_{n\times n}	&	\Id_{n\times n} \\
			-\Id_{n\times n} & 0_{n\times n}
		\end{array}
	\right).
\end{equation*}

We end this section with some miscellaneous remarks.

A state vector $\ket\psi.$ can be identified with a complex function on
configuration space by setting $\psi(q)=\braket q, \psi.$. We will use
both representations interchangeably.

The continuous Weyl operators $w(p,q)=e^{i(p \hat X - q \hat P)}$,
$p,q\in \RR$ fulfill exactly the same composition law as stated in Eq.
(\ref{heisenbergComp}), if $\chi$ is set to $\chi(q)=e^{iq}$ and the
other symbols are interpreted in the obvious way. In fact, Eq.
(\ref{heisenbergComp}) is then equivalent to the fundamental
\emph{Weyl commutation relations} \cite{folland}.  Having this analogy
in mind, $p$ and $q$ will sometimes be called \emph{momentum} and
\emph{position} coordinates respectively.

For future reference, note the two simple relations
\begin{eqnarray}
 	\left( w(p,q) \psi \right) (x) &=& \chi(-2^{-1} pq+px) \psi(x-q),
	\label{weylCoordinates} \\
	\tr w(p,q) &=& d^n \, \delta_{p,0}\delta_{q,0}.
	\label{weylTrace}
\end{eqnarray}

It remains yet to justify the name Weyl \emph{representation}. For
$v\in V, t\in \ZZ_d$, define $w(v,t)=\chi(t)w(v)$. Equation
(\ref{heisenbergComp}) takes on the form 
\begin{equation*}
	w(v_1,t_1)w(v_2,t_2)=w(v_1+v_2,t_1+t_2+2^{-1} [v_1,v_2]).
\end{equation*}
The set $V\times\ZZ_d$, equipped with the above composition law is
called the \emph{Heisenberg group} $H(\ZZ_d^n)$, the Weyl matrices
constituting a unitary representation of $H(\ZZ_d^n)$ \cite{folland}.
This point of view on Weyl operators will be needed only in Appendix
\ref{scCliffordProof}.

\subsection{Clifford group}

The Clifford group is the subset of the unitary operators that map
Weyl operators to multiples of Weyl operators under conjugation:
\begin{equation}\label{cliffordDefinition}
	U w(v) U^\dagger = c(v) w(S(v))
\end{equation}
for some maps $c: V \to \CC$ and $S: V \to V$
\cite{gottesman}. 
The structure of the Clifford group 
is described in the following
theorem
\footnote{
	Note that the ``Clifford group'' which appears in the context of
	quantum information theory \cite{gottesman} has no connection to the
	group by the same name used	e.g.\ in the representation theory of
	$SO(n)$.
}.

Before stating the theorem, we have to comment on a re-appearing
issue: namely that things are more involved if $d$ is not a prime
number. For prime values of $d$, $\ZZ_d$ has the structure of a
\emph{finite algebraic field}, $\ZZ_d^n$ is a \emph{finite vector
space} and most of the intuitions we have about vector spaces continue
to be true.  Among the more severe deficiencies of the general case is
the fact that not every element $a$ of $\ZZ_d$ possesses a
multiplicative inverse modulo $d$. But even if the analogue of a
theorem about vector spaces holds for non-prime values of $d$, it is
often difficult to find a proof in the literature.  Appendices
\ref{scCharacters} and \ref{scGeometric} contain a collection of
statements of this kind.
Less technically inclined readers will not loose much by skipping
these sections.

For the sake of clarity of language, we call functions $f$ on $Q$ which
fulfill $f(\lambda a +b)=\lambda f(a) + f(b)$ \emph{linear},
disregarding the fact that $Q$ might fail to be a linear space.
Similarly, a subset $S$ of $Q$ that is closed under addition and
multiplication by elements of $\ZZ_d$ is referred to as a
\emph{subspace}.  
We define a function $S$ to be
\emph{symplectic} if it is linear and preserves the symplectic form:
$[S\spacedot,S\spacedot]=[\spacedot,\spacedot]$.

\pagebreak
\begin{theorem} \emph{(Structure of the Clifford group)}
	\label{thCliffordStructure}
	\begin{enumerate}
		\item
		For any symplectic $S$, there is a unitary operator
		$\mu(S)$ such that
		\begin{equation*}
			\mu(S)\,w(v)\,\mu(S)^\dagger = w(S\,v).
		\end{equation*}

		\item
		$\mu$ is a \emph{projective representation} of the symplectic
		group, that is
		\begin{equation*}
			\mu(S)\mu(T)=e^{i \phi} \mu(ST)
		\end{equation*}
		for some phase factor $e^{i\phi}$.


		\item
		\label{cliffordDecomposition}
		Up to a phase, any Clifford operation is of the form
		\begin{equation*}
			U=w(a)\mu(S)
		\end{equation*}
		for a suitable $a\in V$ and symplectic $S$.
	\end{enumerate}
\end{theorem}

The representation $\mu$ is called the \emph{Weil} or
\emph{metaplectic} representation \cite{weil,folland}.  Theorem
\ref{thCliffordStructure} is could be called a discrete version of the
celebrated \emph{Stone-von Neumann Theorem} \cite{folland}. Its proof
is not essential for understanding the further argument and has
therefore been moved to Appendix \ref{scCliffordProof}.

Note that a Clifford operation is connected to a vector $a$ and a
linear mapping $S$. This should remind us of a well-known structure on
linear spaces: \emph{affine transformations}. An affine mapping $A$ is
of the form $A(b) = S\,b + a$ where $S$ is an invertible linear
operator and $a$ a vector.  Let us call $A$ symplectic if its linear
part $S$ is.

We will frequently use the 'dot notation' to define functions of one
parameter; for example writing $S\cdot+\,a$ for $A$.

\begin{lemma} \emph{(Clifford group and affine transformations)}
	\label{cliffordAffine}
	The mapping
	\begin{equation*}
		S\cdot+\,a \mapsto w(a)\mu(S)
	\end{equation*}
	is a projective representation of the group of symplectic affine
	transformations.
\end{lemma}

\begin{proof}
	All we need to do is to compare the composition law of the affine
	group
	\begin{eqnarray*}
		(S\cdot+a)\circ(T\cdot+b) &=& S(T\cdot + b)+a \\
		&=& ST \cdot + (Sb +a)
	\end{eqnarray*}
	to the composition law of the representation
	\begin{eqnarray*}
		w(a)\mu(S)\;w(b)\mu(T) 
		&=& w(a)\;\mu(S)w(b)\mu(S)^\dagger\;\mu(S)\mu(T) \\
		&=& w(a) w(Sb) \mu(S)\mu(T) \\
		&\propto& w(Sb+a) \mu(S\,T)
	\end{eqnarray*}
	which proves the assertion. 
\end{proof}

The correspondence established by the last lemma will find a very
tangible manifestation in Section \ref{scWigner}, when we will see
that the Clifford group induces affine transformations of the Wigner
function.

\subsection{Fourier Transforms}
\label{scFourier}

Let $Q=\ZZ_d^n$ and $f: Q\to\CC$ be a complex function on $Q$. 
The Fourier transform of $f$ is
\begin{eqnarray}\label{fourierTrans}
	(\Fou f)(p) =\hat f(p)
	&=& |Q|^{-1/2} \sum_{q\in Q} \bar\chi(pq) f(q).
\end{eqnarray}

In the course of the main proof we will be confronted with Fourier
transforms of functions which are defined only on a subspace of $Q$.
If $d$ is prime, then any subspace of $Q=\ZZ_d^n$ is of the form
$\ZZ_d^{n'}$, for some $n'\leq n$, so no new situation arises. 

For non-prime dimensions, however, subspaces may not be as
well-behaved. Consider as an example $\{0,3,6\}\subset\ZZ_9^1$.  The
set is closed under addition and multiplication, but can clearly not
be written as $\ZZ_9^{n'}$. 

To cope with this problem, we will cast Eq.  (\ref{fourierTrans}) into
a form that is well-defined for functions $f$ on more general spaces.
The construction is presented below.  It can be found in any textbook
on harmonic analysis (e.g.  Ref. \cite{harmonictextbook}).

A \emph{character} of $Q$ is a function $\zeta: Q\to \CC$ such that
$\nolinebreak{\zeta(a+b)}=\zeta(a)\zeta(b)$. Any character of $Q$ is
of the form $\zeta(q)=\bar\chi(xq)$ for an appropriate $x\in Q$ (see
Appendix \ref{scCharacters}). We can hence conceive the Fourier
transformation defined in Eq. (\ref{fourierTrans}) as a function of
the characters of $Q$:
\begin{equation}\label{generalFourierTrans}
	\hat f(\zeta) = |Q|^{-1/2} \sum_q \zeta(q) f(q).
\end{equation}
We denote the set of characters of $Q$ by $Q^*$. With these notions, Eq.
(\ref{generalFourierTrans}) defines a function $Q^* \to \CC$. If, now,
$S$ is any subspace of $Q$ and $f$ a function on $S$, the Fourier
transform
\begin{equation*}
	\hat f: S^* \to S \quad\quad 
	\hat f(\zeta)=|S|^{-1/2} \sum_s \zeta(s) f(s)
\end{equation*}
is well-defined.

For $f: V\to\CC$, we define the \emph{symplectic Fourier
transform} as
\begin{equation}
	(\Fou_S f)(a)
	= |V|^{-1/2} \sum_{b\in V} \bar\chi([a,b]) f(b).
\end{equation}

Finally, take a note that the normalization in Eqs.
(\ref{fourierTrans}) and (\ref{generalFourierTrans}) has been
chosen in such a way that \emph{Parzeval's Theorem} $||f||=||\hat f||$
holds, where $||f||^2=\sum_q |f(q)|^2$.

\subsection{Definition and properties of the Wigner function}
\label{scWigner}

Employing Eq. (\ref{weylTrace}) in conjunction with the composition
law Eq. (\ref{heisenbergComp}), one finds that the Weyl operators
$\{w(p,q)\}$ form an orthonormal basis in the space of operators on
$\H$ with respect to the trace scalar product
$d^{-n}\,\tr(\spacedot^\dagger\spacedot)$. The \emph{characteristic
function} $\Xi_\rho$ of an operator $\rho$ is given by its expansion
coefficients with respect to the Weyl basis:
\begin{equation}
	\Xi_\rho(\xi,x)=d^{-n}\, \tr(w(\xi,x)^\dagger \rho).
\end{equation}

We mentioned in the introduction that the continuous Wigner function
is the symplectic Fourier transform of the characteristic function
\cite{wallsmilburn,folland}. The two latter concepts have been defined
for finite-dimensional systems in the preceding paragraphs. We can now
state, in complete analogy to the continuous case:

\begin{definition}\label{dfWigner}
	\emph{(Wigner function)}
	Let $d$ be odd, $Q=\ZZ_d^n$ for some $n$. Let $V, \H$ be as usual
	and let $\rho$ be a quantum state on $\H$.

	The \emph{Wigner function} $W_\rho$ associated with $\rho$ is the
	symplectic Fourier transformation of the characteristic function
	$\Xi_\rho$.
\end{definition}

An explicit calculation yields, for all $a\in V$,
\begin{eqnarray}
	\left(\Fou_S\,\Xi_\rho\right)(a) 
	&=& d^{-2n} \, \sum_{b\in V} \bar\chi([a,b]) \tr(w(b)^\dagger \rho) 
	\nonumber \\
	&=& d^{-n} \tr( \left(d^{-n}\, \sum_{b} \bar\chi([a,b]) w(b)^\dagger\right)
	\rho) \nonumber \\
	&=:& d^{-n} \tr ( A(a) \rho ), \label{pspo}
\end{eqnarray}
where we have implicitly defined the \emph{phase space point
operator} $A(a)$ \cite{wootters}.

Theorem \ref{thWignerProperties} lists a selection of properties of
the Wigner function. For a more thorough discussion, the reader is
deferred to Refs. \cite{diploma,vourdas}.

\begin{theorem}\label{thWignerProperties}
	\emph{(Properties of the Wigner function)}

	\begin{enumerate}
		\label{wignerOrthonormal}
		\item
		The phase space point operators have unit trace and form an
		orthonormal basis in the space of Hermitian operators on $\H$.
		Hence the Wigner function of an Hermitian operator is real, and
		further, the \emph{overlap}
		\begin{equation*}
			d^{-n}\, \tr(\rho\sigma)
			= 
			\sum_{v\in V} W_\rho(v) W_\sigma(v),
		\end{equation*}
		and \emph{normalization} relations 
		\begin{equation*}
			\sum_v W_\rho(v) = \tr \rho
		\end{equation*}
		hold.

		\item\label{wignerPureState}
		For a pure state $\psi$, the Wigner function 
		$W_\psi:=W_{\ket\psi.\bra\psi.}$ equals
		\begin{eqnarray*}
			&&W_\psi(p,q) = \\
			&&d^{-n} \sum_{\xi\in Q} \bar\chi(\xi p) 
			\bar\psi(q-2^{-1} \xi)\psi(q+2^{-1} \xi).
		\end{eqnarray*}

		\item\label{wignerMarginal}
		When computing marginal probabilities, the Wigner function behaves
		like a classical probability distribution:
		\begin{equation*}
			\sum_{p\in Q} W_\psi(p,q)=|\psi(q)|^2.
		\end{equation*}

		\item\label{wignerFactor}
		The multi-particle phase space point operators factor: 
		\begin{equation*}
			A(p_1, \dots, p_n, q_1, \dots, q_n) 
			= \bigotimes_i^n A^{(i)}(p_i, q_i)
		\end{equation*}
		(and hence so does the Wigner function).

		\item\label{wignerParity}
		It holds that $A(0)\ket q.=\ket -q.$. In other words, the phase
		space point operator at the origin equals the \emph{parity
		operator}.

		\item\label{wignerMoyal}
		The Wigner function $W_{\rho\,\sigma}$ of an operator product  is
		given by the \emph{$\star$-product} (also known as the \emph{Groenewold} or
		\emph{Moyal product} \cite{groenewold}):
		\begin{eqnarray*}
			&& W_{\rho\,\sigma}(u)=(W_\rho \star W_\sigma)(u) \\
			&:=& d^{-n} \sum_{v,w} W_\rho(u+v) \, W_\sigma(u+w)
			\bar\chi([v,w]).
		\end{eqnarray*}
	\end{enumerate}
\end{theorem}

\begin{proof}
	The proofs are all straight-forward; we give only hints on how to
	conduct them. It will be essential to recall  the well-known
	relation
	\begin{equation}\label{characterSums}
		\sum_{x\in \ZZ_d^n} \chi(x y)= d^n\,\delta_{y,0},
	\end{equation}
	for all $y \in \ZZ_d^n$.

	Indeed, the first claim can be proven by using Eq.
	(\ref{characterSums}) together with the definition of the
	phase space point operators Eq. (\ref{pspo}). Employ Definition
	\ref{dfWigner} and Eq. (\ref{characterSums}) to establish the second
	assertion, which in turn implies the third one. Theorem
	\ref{thWignerProperties}.\ref{wignerFactor} makes use of the fact
	that 
	$
		\bar\chi(p q)=\prod_i \bar\chi(p_i q_i);
	$
	see also Section \ref{scPrimePower} for a very similar and more
	explicit calculation. The validity of the fifth statement is best
	shown using Eqs. (\ref{weylCoordinates}), (\ref{characterSums}).
	
	Let us lastly turn to Claim \ref{wignerMoyal}.  We have noted that
	the phase space point operators form an orthonormal system. Hence we
	can expand an operator $\rho$ in terms of its Wigner function as
	$\rho = \sum_v W_\rho(v) A(v)$. Substituting $\rho$ and $\sigma$ by
	their respective expansions in
	$W_{\rho \sigma}(v) = d^{-n} \tr(A(v) \rho \sigma)$ yields the
	desired formula with the help of Lemma \ref{lmAProps}.
\end{proof}

The following statement will be vital to the proof of the main
theorem. It assigns an elegant geometric interpretation to the
Clifford group.

\begin{theorem}\label{thWignerCovariance}
	\emph{(Clifford Covariance)}
	Let $U=w(a)\mu(S)$ be a Clifford operation. Let $\rho':=U\rho
	U^\dagger$ for some Hermitian operator $\rho$. The Wigner function
	is \emph{covariant} in the sense that
	\begin{equation*}
		W_{\rho}(v) = W_{\rho'}(S\,v+a).
	\end{equation*}
\end{theorem}

\begin{proof}
	We compute the action of the Clifford group on the phase space point
	operators.
	\begin{eqnarray*}
		&&w(a)\mu(S)\,A(b)\,\mu(S)^\dagger w(a)^\dagger\\
		&=&
		d^{-n} \sum_{v\in V} \bar\chi([b,v])
		w(a)\mu(S)\,w(v)\,\mu(S)^\dagger w(a)^\dagger \\
		&=&
		d^{-n} \sum_{v} \bar\chi([b,v])
		w(a)\,w(S\,v)\,w(a)^\dagger \\
		&=& d^{-n} \sum_{v} \bar\chi([b,v])\chi([a,S\,v]) w(S\,v) \\
		&=& d^{-n} \sum_{v':=S\,v} \bar\chi([b,S^{-1}v'])\bar\chi([a,v']) w(v') \\
		&=& d^{-n} \sum_{v'} \bar\chi([S\,b+a,v']) w(v') 
		= A(Sb+a).
	\end{eqnarray*}
	The claim follows by use of Eq. (\ref{pspo}).
\end{proof}

Our definition of the discrete Wigner function coincides with the ones
used in Refs.  \cite{oldWootters,vourdas,villegar,diploma}. It is
further equal to Leonhardt's version \cite{leonhardt}, up to a
permutation of points in phase space; it corresponds to
choice (a) in Ref. \cite{klimov} and lastly to $G=\ZZ_d^n$ in Ref.
\cite{chaturvedi}.  One can show that $W$, as defined
here, fulfills the axioms of Ref.  \cite{wootters} which had been laid
out in Section \ref{scPrevious}. Put differently, it is an element of
the set of generalized Wigner functions. Gibbons \emph{et.  al.}
remarked in Ref. \cite{wootters} that among the generalized Wigner
functions, some stand out by their high degree of symmetry. In our
language, this symmetry is an incarnation of the Clifford covariance
established in Theorem \ref{thWignerCovariance}. Naturally, it is now
interesting to ask how much freedom is left in the definition of a
Wigner function, once one requires Clifford covariance to hold. We
show in Appendix \ref{scUniqueness} that the definition used here is
virtually unique in that regard.

\subsection{Stabilizer States}

Using the composition law of the Heisenberg group Eq.
(\ref{heisenbergComp}), it is easy to see that two Weyl operators
$w(v_1), w(v_2)$ commute if and only if $[v_1, v_2]=0$. Now consider
the image of an entire subspace $M$ under the Weyl representation $w$. 
The set 
\begin{equation*}
	w(M)=\{w(m)|m\in M\}
\end{equation*}
consists of mutually commuting operators if and only if the symplectic
form vanishes on $M$:
\begin{equation*}
	[m_1,m_2]=0, \quad \text{for all }m_i\in M.
\end{equation*}
Spaces of that kind are called \emph{isotropic}.  Clearly, if $M$ is
isotropic, then the operators $w(M)$ can be simultaneously
diagonalized. We will see that if $|M|=d^n$, the eigenspaces become
non-degenerate and can thus be used to single out state vectors in the
Hilbert space. A subspace $M$ of $V$ is said to be \emph{maximally
isotropic} if its cardinality equals $d^n$.  See Appendix
\ref{scCharacters} for a justification of that nomenclature.

\begin{lemma}\label{lmStabilizers}
	\emph{(Stabilizer States)}
	Let $M$ be a maximally isotropic subspace of $V$. Let $v \in V$.
	Up to a global phase, there is a unique state vector $\ket M, v.$
	that fulfills the eigenvalue equations
	\begin{equation*}
		\chi([v,m])w(m) \,\ket M, v. = \ket M, v.
	\end{equation*}
	for all $m\in M$.
\end{lemma}

\begin{proof}
	Existence:
		It is elementary to check that
		\begin{equation}\label{stabProjector}
			|M|^{-1} \sum_{m\in M} \chi([v,m]) w(m)
		\end{equation}
		is a rank one projection operator fulfilling the eigenvalue
		equations.

	Uniqueness:
		According to Appendix \ref{scCharacters}, there are $p^n$
		characters of $M$, each giving rise to a distinct projection
		operator as defined in the last paragraph. Two distinct operators
		of that kind are mutually orthogonal, because they belong to
		different eigenvalues of at least one of the Weyl operators.  But
		$\dim\H=|Q|=p^n$ and thus there is no space for more than
		one-dimensional solutions to the given set of equations.
\end{proof}

The state vector $\ket M,v.$ is called the \emph{stabilizer state}
associated to $M$ and $v$. For obvious reasons, one refers to the set
of operators $\{\chi([v,m])\,w(m)|m\in M\}$ as the \emph{stabilizer}
of $\ket M,v.$. Due to the isotropicity of $M$, the stabilizer is
closed under multiplication and thus constitutes a group.
Occasionally, we write $\ket M.$ for $\ket M, 0.$. To specify a
stabilizer state, we need to specify a maximally isotropic space $M$.
This is best done by giving a basis $\{m_1, \dots, m_k\}$ of $M$.  It
is convenient to assemble the basis vectors as the columns of a
$2n\times k$-matrix, which is generally referred to as the
\emph{generator matrix}. As the choice of a basis is non-unique, so is
the form of the generator matrix.

A stabilizer state $\ket M.$ is a \emph{graph state} if it possesses a
generator matrix of the form 
\begin{equation}\label{graphState} 
	\left(
		\begin{array}{c} 
			\vartheta \\ 
			\Id_{n\times n}
		\end{array} 
	\right),
\end{equation} 
where $\vartheta$ is a symmetric $n\times n$-matrix \cite{jensGraph}. The
designation stems from the fact that $\vartheta$ can be interpreted as
the adjacency matrix of a graph. Many properties of $\ket M.$ are
describable in terms of that graph alone \cite{jensGraph}.
Some authors require the diagonal elements ${\vartheta^i}_i$ to vanish
(equivalently, no vertex of the graph should be linked to itself), but
we will not impose that restriction. Note that there exist
considerably more general definitions of graph states
\cite{schlinge}.

Obviously, we will be concerned with Wigner functions of stabilizer
states. Lemma \ref{lmStabilizerWigner} clarifies their structure.

\begin{lemma}\label{lmStabilizerWigner}
	\emph{(Wigner functions of stabilizer states)}
	The Wigner function of a stabilizer state $\ket M, v.$ is the
	\emph{indicator function} on $M+v$. More precisely,
	\begin{equation*}
		W_{\ket M, v.}(a) = 
		\frac1{d^n} \delta_{M+v}(a) =
		\frac1{d^n} 
			\left\{
				\begin{array}{ll}
					1\quad	&	a \in M+v \\
					0 			& \mathrm{else}.
				\end{array}
			\right.
	\end{equation*}
\end{lemma}

\begin{proof}
	The representation given in Eq. (\ref{stabProjector}) of $\ket M,v.$
	determines the characteristic function
	\begin{equation*}
		\Xi_{\ket M,v.}(b) = d^{-n}\,\chi([v,b]) \delta_M(b).
	\end{equation*}
	We compute the symplectic Fourier transformation:
	\begin{eqnarray*}
		\big(\Fou_S\,\Xi_{\ket M,v.}\big)(a)
		&=& d^{-2n} \sum_{b\in V} \bar\chi([a,b])\chi([v,b])\delta_M(b) \\
		&=& d^{-2n} \sum_{b\in M} \bar\chi([a-v,b]) \\
		&=& d^{-n}\, \delta_{M^\bot}(a-v).
	\end{eqnarray*}
	Where 
	\begin{equation*} 
		M^\bot=\{v\in V|[m,v]=0 \text{ for all } m\in M\} 
	\end{equation*} 
	is the \emph{symplectic complement} of $M$ in $V$. But $M$ is a
	maximally isotropic space and hence $M=M^\bot$ (see Appendix
	\ref{scCharacters}).
\end{proof}

In particular we know now that the Wigner function of stabilizer
states is non-negative. The next sections are devoted to the proof of
the converse. 

\section{Discrete Hudson's Theorem}
\label{scMain}

\subsection{Bochner's Theorem}

Define the \emph{self correlation function} 
\begin{equation*}
	K_\psi(q,x)=\psi(q+2^{-1}x) \bar\psi(q-2^{-1}x)
\end{equation*}
and note that the Wigner function fulfills
\begin{eqnarray}\label{wignerCorr}
	W(p,q)
	&=&\frac1{d^n} \sum_{x\in Q} \bar\chi(p x) K_\psi(q,x).
\end{eqnarray}
Fix a $q_0\in Q$. Designating the function $p\mapsto W(p,q_0)$ by
$W(\spacedot,q_0)$, Eq. (\ref{wignerCorr}) says that $W(\spacedot,q_0)$ is
the Fourier transform of $K(q_0,\spacedot)$.  Therefore, $W$ is
non-negative if and only if the $d^n$ functions $K(q_0,\spacedot)$ have
non-negative Fourier transforms.

In harmonic analysis, the set of functions with non-negative Fourier
transforms is characterized via a theorem due to Bochner. 
It is usually proven either in the
context of Fourier analysis on the real line or else, in full
generality, for harmonic analysis on -- not necessarily abelian -- locally
compact groups. While the former statement is not general enough for
our purpose, the latter is not easily accessible. However, it turns
out that in the discrete abelian setting an elementary proof can be
given. It is stated in the next theorem, along with a variation
for subsequent use.

\begin{theorem}\label{thBochner}
	\emph{(Variations of Bochner's Theorem)}
	Let $M$ be a subspace of $Q$.  Let $f: M\to \CC$. It holds that
	\begin{enumerate}
		\item\label{bochnerOne} 
		The Fourier transform of $f$ is non-negative if and only if the
		matrix
		\begin{equation*}
			{A^x}_q=f(x-q)\quad\quad (x,q\in M)
		\end{equation*}
		is positive semi-definite.

		\item\label{bochnerTwo}
		The Fourier transform of $f$ has constant modulus (i.e. $|\hat
		f(x)|=\const$) if and only if	$f$ is orthogonal to its
		translations:
		\begin{eqnarray*}
			\tuple f, \hat x(q) f. = \sum_{x\in M} \bar f(x) f(x-q) = 0
		\end{eqnarray*}
		for all non-zero $q\in M$.
	\end{enumerate}
\end{theorem}

\begin{proof}
	The following computation is a variant of a well-known fact
	concerning circulant matrices.  We claim that any character $\zeta$
	of $M$ is an eigenvector of $A$ with eigenvalue
	$\lambda=|M|^{-1/2}\,\hat f(\zeta)$. Indeed, plugging in the
	definitions yields
	\begin{eqnarray*}
		(A\, \zeta)(x)   
		&=& \sum_{q} {A^x}_q\, \zeta(q) \\
		&=& \sum_{q} f(x-q) \zeta(q) \\
		&=& \sum_{q} f(q) \bar\zeta(q)\, \zeta(x) \\
		&=& \sqrt{|M|}\hat f(\zeta)\, \zeta(x). 
	\end{eqnarray*}
	There exist $|M|$ characters and thus equally many eigenvectors of
	$A$. Therefore, $A$ can diagonalized.  All its eigenvalues are
	non-negative if and only if $\hat f$ is non-negative.

	By the same argument, $A$ is proportional to a unitary matrix if and
	only if $|\hat f(q)|$ is constant. But a matrix is unitary if and
	only if its rows form an ortho-normal set of vectors.
\end{proof}

From here, the proof proceeds in two steps. Section
\ref{scHudsonModulus} harvests Theorem
\ref{thBochner}.\ref{bochnerOne} to gain information on the pointwise
modulus $|\psi(q)|$ of a vector with non-negative Wigner function.
Building on these finding, we will analyze the properties of such
Wigner functions in Section \ref{scHudsonWigner}.

\subsection{Supports and Moduli}
\label{scHudsonModulus}

\begin{lemma} \label{lmIneq} 
	\emph{(Modulus Inequality)}
	Let $\psi$ be a state vector with non-negative Wigner function.

	It holds that
	\begin{equation*}
		|\psi(q)|^2 \geq |\psi(q-x)|\,|\psi(q+x)|
	\end{equation*}
	for all $q,x \in Q$.
\end{lemma}

\begin{proof}
	Fix a $q \in Q$.  As $W_\psi$ is non-negative, so is the Fourier
	transform of $K_\psi(q,\spacedot)$. Bochner's Theorem implies that the
	matrix ${A^x}_y = K(x-y,q)$ is positive semi-definite which in turn
	implies that all principal sub-matrices are psd. In particular the
	determinant of the $2\times 2$ principal sub-matrix 
	\begin{eqnarray*}
		&&
		\left(
		\begin{array}{cc}
			K_\psi(q,0) & K_\psi(q,2x) \\
			K_\psi(q, -2x) & K_\psi(q,0) 
		\end{array}
		\right) \\
		&=&
		\left(
		\begin{array}{cc}
			|\psi(q)|^2	&	\psi(q+x)\bar\psi(q-x) \\
			\bar\psi(q+x)\psi(q-x) & |\psi(q)|^2 
		\end{array}
		\right)
	\end{eqnarray*}
	must be non-negative. But this means
	\begin{eqnarray*}
		|\psi(q)|^4 - |\bar\psi(q+x)	\psi(q-x)|^2 \geq 0,
	\end{eqnarray*}
	which proves the theorem.
\end{proof}

We will call the set $\supp\psi$ of points where a state-vector is
non-zero its \emph{support}.  $S=\supp\psi$ has the property to
contain the \emph{midpoint} of any two of its elements. Indeed, if $a, b
\in S$, then setting $q=2^{-1} (a+b)$ and $x=2^{-1}(a-b)$ in the
Modulus Inequality shows that
\begin{equation*}
	|\psi(2^{-1} (a+b))| \geq |\psi(a)|\,|\psi(b)| > 0,
\end{equation*}
hence $2^{-1}(a+b) \in S$. Let us refer to sets possessing this quality
as being \emph{balanced}.

The following lemma clarifies the structure of balanced sets.  Recall
that a subset $A$ of $V$ is \emph{affine} if $A=M+v$ for a subspace
$M$ and some vector $v$. An affine space is a subspace if and only if
it contains the origin $0$.  

\begin{lemma}\label{lmBalancedSets}
	\emph{(Balanced sets)}
	A subset $S$ of $Q$ is balanced if and only if $S$ is an affine
	space.
\end{lemma}

\begin{proof}
	We show the 'only if' part, the other one being simple.  
	
	As both the characterizations of balancedness and affinity are
	invariant under translation, there is no loss of generality in
	assuming that $0\in S$. We have to establish that $S$ is closed
	under both addition and scalar multiplication.
	
	Let $a\in S$.  We claim that
	\begin{equation}\label{balancedProp}
		2^{-l}\lambda \,a \in S
	\end{equation}
	for all $l\in \NN$ and $\lambda\leq2^l$. The proof is by induction on $l$.
	Suppose Eq.  (\ref{balancedProp}) holds for some $l$. If
	$\lambda \leq 2^{l+1}$ is even, then $2^{-l-1}\lambda\,a=
	2^{-l}(\lambda/2)b\in S$. Else, 
	\begin{eqnarray*}
		2^{-l-1} \lambda\, a 
		= 
		2^{-1}\big( 2^{-l} \frac{\lambda-1}2\, a + 2^{-l} \frac{\lambda+1}2\,a \big)
		\in S,
	\end{eqnarray*} 
	which shows the validity of Eq. (\ref{balancedProp}).

	There exists an integer $l>d$ such that $2^l = 1 \mod d$.  Indeed,
	by Euler's Theorem, $2^{\phi(d)}=1 \mod d$,  where $\phi$ is
	Euler's totient function. So $l=d \phi(d)$ satisfies the
	requirements.
	Inserting $l$ into Eq. (\ref{balancedProp}), we conclude that
	$\lambda\, a \in S$ for all $\lambda \leq 2^d$. Thus certainly
	$\lambda\,a \in S$ for all $\lambda \in \ZZ_d$ and we have proved
	closure under scalar multiplication. 
	
	If $a,b\in S$ then, by the last paragraph $2a,2b \in S$ and hence
	$2^{-1}(2a+2b)\in S$, establishing closure of $S$ under addition.
\end{proof}

\begin{lemma}\label{lmModulus}
	\emph{(Constant Modulus)} 
	Let $\psi$ be a state vector with non-negative Wigner function.
	Then $|\psi(\spacedot)|$ is constant on the support of $\psi$.
\end{lemma}

\begin{proof}
	Pick two points $x, q \in \supp\psi$ and suppose $|\psi(q)|>|\psi(x)|$.
	
	Letting $z=x-q$, the assumption reads $|\psi(q)|>|\psi(q+z)|$.  The
	Modulus Inequality, centered at $q+z$, gives
	\begin{eqnarray}\label{lcRoot}
		|\psi(q+z)|^2 
		&\geq& |\psi(q)|\,|\psi(q+2z)|.
	\end{eqnarray}
	As $\supp\psi$ is affine, we know that $\psi(q+kz)\neq0$ for all
	$k\in \ZZ_d$. Hence Eq. (\ref{lcRoot}), together with the assumption
	implies
	\begin{eqnarray*}
		&& |\psi(q+z)|^2 > |\psi(q+z)| \, |\psi(q+2z)| \\
		&\Leftrightarrow&  |\psi(q+z)| > |\psi(q+2z)|.
	\end{eqnarray*}
 	By inducting on this scheme, we arrive at
	\begin{equation*}
		|\psi(q)| > |\psi(q+z)| > |\psi(q+2 z)| > \cdots
	\end{equation*}
	and therefore $|\psi(q)|>|\psi(q+dz)|=|\psi(q)|$, 
	which is a contradiction. 

	Thus $|\psi(q)|\leq|\psi(x)|$. Swapping the roles of $x$ and $q$
	proves that equality must hold.
\end{proof}

At this point, we have full knowledge of the pointwise \emph{modulus}
of a state vector with non-negative Wigner function. The \emph{phases}
of $\psi(\spacedot)$ are, however, completely unknown. The section to come
addresses this problem indirectly, by studying non-negative Wigner
functions. 

\subsection{Non-negative Wigner functions}
\label{scHudsonWigner}

To motivate the following, assume for a moment that $\psi$ has a
non-negative Wigner function and further, that $\psi(q)\neq 0$ for all
$q$.
Choose a $q_0\in Q$ and consider the function $W(\spacedot,q_0)$.
Lemma \ref{lmModulus} implies that $K_\psi(q_0,\spacedot)$ has constant
modulus and hence -- by Theorem \ref{thBochner}.\ref{bochnerTwo} --
$W(\spacedot,q_0)$ must be orthogonal to its translations.  Clearly, a
non-negative function possesses this property if and only if it is
supported on at most a single point.

There hence exists a $p_0\in Q$ such that
$W(p,q_0)\propto\delta_{p,p_0}$. This observation starkly reduces the
possible forms of positive Wigner functions; it will be generalized to
state vectors with arbitrary support in the next lemma.

\begin{lemma}\label{lmWignerConstant}
	Let $\psi$ be a state vector. If $W_\psi$ is non-negative, then it
	is of the form
	\begin{equation*}
		W_\psi(v) = d^{-n}\,\delta_T(v)
	\end{equation*}
	where $T\subset V$ is a set of cardinality $d^n$. 
	
	What is more, if $0\in T$, then the set of elements of $T$ with
	vanishing position coordinates
	\begin{equation*}
		\{(p,0)\in T \,|\,  p \in Q\}
	\end{equation*}
	is a subspace of $V$.
\end{lemma}

\begin{proof}
	Let $S=\supp\psi$. Again, we may assume that $S$ is a subspace of
	$Q$, for else we replace $\psi$ by $w(-s)\psi$ for some $s\in S$.
	It follows that $\supp K_\psi = S\times S$. Indeed, 
	\begin{eqnarray*}
		K_\psi(q,x) \neq 0 
		&\Leftrightarrow& q\pm 2^{-1}x \in S \\
		&\Leftrightarrow& q \in S \wedge x \in S.
	\end{eqnarray*}
	
	Denote by $S^\bot=\{q\in Q| sq =0 \text{ for all } s\in S\}$ the
	orthogonal complement of $S$
	\footnote{
		For subsets $S$ of $Q$, $S^\bot$ denotes the \emph{orthogonal}
		complement, while for subsets $S$ of $V$ the same symbol refers to
		the \emph{symplectic} complement. This notation is natural, as for
		both $Q$ and $V$ only one respective inner product has been
		defined. 
	}.  
	We will adopt the common notation $[p]=p+S^\bot$ for cosets of
	$S^\bot$. It should be clear that $[p]$ is nothing other but the
	\emph{affine space} with directional vector space given by $S^\bot$
	and base vector $p$. The set $S^*$ of characters of $S$ can be
	identified with $Q/S^\bot$. Certainly, $s \mapsto \chi(ps)$ defines
	a character of $S$ for every $p\in Q$.  Further,
	$\chi(ps)=\chi(p's)$ for all $s\in S$ if and only if $p-p'\in
	S^\bot$. That indeed all elements of $S^*$ can be obtained this way
	is shown in Corollary \ref{crComplements}.

	Define $K'_\psi$ to be the restriction of $K_\psi$ to its support
	$S\times S$. For the rest of the proof, we fix a $q_0\in S$. Now consider 
	\begin{eqnarray*}
		W(p,q_0)
		&=& d^{-n} \sum_{x\in Q} \bar\chi(px) K(q_0,x) \\
		&=& d^{-n} \sum_{x\in S} \bar\chi(px) K'(q_0,x).
	\end{eqnarray*}
	Viewed as a function in $p$, $W(p,q_0)$ has constant values on
	cosets of $S^\bot$.  Therefore, 
	\begin{eqnarray}\label{reducedTransform}
		W'([p],q):=d^n|S|^{-1/2}\,W(p,q)
	\end{eqnarray}
	is a well-defined function on $S^*$. The considerations of the
	previous paragraph allow us to identify $W'([\spacedot],q_0)$ as the
	Fourier transform of $K'(q_0,\cdot)$. 

	We can now repeat the argumentation presented just before the
	current lemma. Indeed, the modulus of $K'(q_0,[\spacedot])$ is
	constant and $W'$ is non-negative. Furthermore, by definition of
	$q_0$, $K'(q_0,[\spacedot])$ is non-zero and we may thus conclude that
	$p\mapsto W'([p],q_0)$ is supported on exactly one coset $[p_0]$.

	Normalization of $\psi$ implies $|\psi(\spacedot)|=|S|^{-1/2}$. 
	Hence $|K'_\psi(q_0,\spacedot)|=|S|^{-1}$  and
	\begin{equation*}
		||K'_\psi(q_0,\spacedot)||^2=\sum_x |K'_\psi(q_0,x)|^2=|S|^{-1}.
	\end{equation*}
	By Parzeval's Theorem, $||W'([\spacedot],q_0)||^2=|S|^{-1}$ as well.
	It follows that $W'([p_0],q_0)=|S|^{-1/2}$.
	
	Inverting Eq. (\ref{reducedTransform}) gives
	\begin{equation}\label{wignerIndicator}
		W(p,q)= 
		d^{-n}\,
		\left\{ 
			\begin{array}{ll}
				1	\quad	&	[p]=[p_0] \\
				0				& \text{else}
			\end{array}
		\right.
	\end{equation}
	which proves the first claim of the lemma. The cardinality of $T$ is
	fixed by the normalization of the Wigner function (Theorem
	\ref{thWignerProperties}.\ref{wignerOrthonormal}).

	Now suppose $W(0,0)=W'([0],0)\neq 0$. Clearly, then $W(p,0)$ is
	non-zero if and only if $p\in [0]\Leftrightarrow p \in S^\bot$.
	The last assertion of the lemma follows, since $S^\bot$ is a
	subspace of $Q$.
\end{proof}

So a non-negative Wigner function is the indicator functions of some
set $T$. This finding is compatible with Lemma
\ref{lmStabilizerWigner}, which describes the structure of Wigner
functions of stabilizer states. The next two lemmas verify that $T$
has indeed all the properties of the sets that appear in Lemma
\ref{lmStabilizerWigner}.

\begin{lemma}
	Let $\psi$ be a state vector. If $W_\psi$ is of the form
	\begin{equation*}
		W_\psi(v) = d^{-n}\,\delta_T(v),
	\end{equation*}
	then $T$ is an affine space.
\end{lemma}

\begin{proof}
	The proof proceeds similar to the one of Lemma \ref{lmBalancedSets}. 
	There is no loss of generality in assuming that $0\in T$. 
	
	First, we show that $T$ is closed under scalar multiplication. To
	this end, pick a point $a\in T$. There exists a symplectic mapping
	$S$ that sends $a$ to a vector $a'$ of the form $(a_p',0)$ where
	$a_p'\in Q$ (see Appendix \ref{scGeometric}).  The set $T'=S\,T$ is
	the support of the Wigner function of $\mu(S)\psi$. By the second
	assertion of Lemma \ref{lmWignerConstant}, $\lambda a' \in S\,T$ for
	every $\lambda\in\ZZ_d$. Hence $S^{-1} (\lambda a') = \lambda a \in
	T$.
	
	Turning to closedness under addition, let $a,b\in T$. By the last
	paragraph, $2a, 2b \in T$. Arguing as before, note that the set
	$T-2a$ is the support of the Wigner function of $w(-2a)\psi$ and
	thus closed under multiplication. As $2b-2a\in T-2a$, we know that
	$b-a\in T-2a$ and hence $b+a\in T$.  
\end{proof}

\begin{lemma}
	Let $\psi$ be a state vector such that $W_\psi$ is of the form
	\begin{equation*}
		W_\psi(v) = d^{-n} \delta_T(v).
	\end{equation*}
	If $T$ is a subspace,
	then it is isotropic.
\end{lemma}

\begin{proof}
	The vector $\psi$ describes a pure state, hence $W_\psi \star
	W_\psi=W_\psi$ (recall the Moyal product, introduced in Theorem
	\ref{thWignerProperties}).  Let $u\in T$. Plugging in the
	definitions gives
	\begin{eqnarray*}
		&&W_\psi \star W_\psi (u)\\
		=&& d^{-n} 
		\sum_{v,w\in V} W_\psi(u+v) \, W_\psi(u+w) \bar\chi([v,w])\\
		=&& d^{-3n} \sum_{v,w\in T} \bar\chi([v,w]).
	\end{eqnarray*}
	Note that $\sum_{w\in T}\bar\chi([v,w])\leq|T|=d^n$ with equality if
	and only if $[v,w]=0$ for all $w$. Hence
	\begin{eqnarray*}
		W_\psi \star W_\psi(u) 
		\leq d^{-n} = W_\psi(u).
	\end{eqnarray*}
	For the left-hand and the right-hand side to be equal, $T$ must be
	isotropic.
\end{proof}
		
Therefore $T$, as defined above, is of the form $T=M+v$ where $M$ is
an isotropic space of cardinality $d^n$. But then, $W_\psi$ is the
Wigner function of a stabilizer state, by Lemma
\ref{lmStabilizerWigner}. We have proven:

\begin{theorem}\emph{(Main Theorem)}
	Let $\psi\in L^2(\ZZ_d^n)$ be a state vector. If the Wigner function
	of $\psi$ is non-negative, then $\psi$ is a stabilizer state.
\end{theorem}

\section{Discrete Gaussians}
\label{scDiscreteGaussians}

It has long been realized that the coefficients of stabilizer state
vectors are described by quadratic forms.  However, the current
literature either neglects the non-prime case (Refs.
\cite{schlinge,grassl,dehaene}) or is less explicit (Ref.
\cite{erik}) than the following lemma in showing the tight relation
between Gaussian states and stabilizer states.

We will concentrate on stabilizer states with full support. This
constitutes only a modest restriction of generality. Indeed, let
$\psi$ be a general stabilizer state, let $Q':=\supp \psi$. Let us for
the sake of simplicity assume that $d$ is prime and $Q'$ is a subspace
of $Q$. The restriction of the coordinate function $\psi(q)$ to $Q'$
can be thought of as defining a vector $\psi'$ of a quantum state of
an $n':=\dim Q'$ particle system. It is now possible to check that
$\psi'$ is a stabilizer state. In this way any stabilizer state can be
viewed as one with full support, possibly on a smaller system. We
will, however, not take the time to make this construction precise nor
will we rely on it in this paper.

\begin{lemma}
	Let $\psi$ be a state vector. The following statements are
	equivalent.
	\begin{enumerate}
		\item\label{one}
		$\psi$ is a stabilizer state and $\psi(q)\neq 0$ for all $q\in Q$.

		\item\label{two}
		Up to the action of a Weyl operator, $\psi$ is a graph state.

		\item\label{three}
		There exists a symmetric $n\times n$-matrix $\theta$ and an $x\in
		Q$ such that
		\begin{eqnarray*}
			\psi(q) = \omega^{q\theta q+x q}.
		\end{eqnarray*}
	\end{enumerate}
\end{lemma}

\begin{proof}
	\emph{(\ref{one} $\Rightarrow$ \ref{two}).}
	By assumption $\ket\psi.=\ket M, v.$ for some maximal isotropic
	space $M$ and a vector $v$. We claim that there is no non-zero $p\in
	Q$ such that $(p,0)\in M$.

	For suppose there exists such a $p$. Then
	\begin{eqnarray*}
		\bra q. w(p,0)\ket M.=\chi(-pq)\braket q, M..
	\end{eqnarray*}
	On the other hand,
	\begin{eqnarray*}
		\bra q. w(p,0)\ket M.=\bar\chi([v,(p,0)])\,\braket q, M.,
	\end{eqnarray*}
	by the definition of $\ket M, v.$. Hence $\supp \ket M.$ must be
	contained within a hyper-surface of $Q$ specified by $pq=\const$,
	which contradicts the assumption that $\supp\psi=Q$.

	There are $d^n$ elements in $M$. By the last paragraph, no two of
	them have the same position coordinates. As there exist only
	$d^n=|Q|$ possible choices for the position coordinates, one can
	find for every $q\in Q$ a $p\in Q$ such that $(p,q)\in M$. Let
	$e_1,\dots,e_n$ denote the canonical basis of $\ZZ_d^n$. Choose
	$m_1,\dots,m_n\in M$ such that the position part of $m_i$ equals
	$e_i$. The span of $\{m_i\}_{i=1,\dots,n}$ has clearly cardinality
	$d^n$, so we have found a basis of $M$. By construction, the
	generator matrix composed of these basis vectors has the form shown
	in Eq. (\ref{graphState}) with some $n\times n$-matrix $\theta$. It
	is not hard to see that $M$ is isotropic if and only if $\theta$ is
	symmetric, establishing that $\ket M.$ is a graph state. Theorem
	\ref{thWignerCovariance}  and Lemma \ref{lmStabilizerWigner} show
	that $w(v)\ket M.=\ket M,v.=\ket\psi.$.

	\emph{(\ref{two} $\Rightarrow$ \ref{three}).}
	Let $M$ be an isotropic space which possesses a generator matrix of
	the form given in Eq. (\ref{graphState}). Let
	$m_i=(\vartheta_i,e_i)$ be the $i$th column of that matrix.
	We need to establish the existence of a symmetric matrix $\theta$
	and an $x\in Q$ such that 
	\begin{equation*}
		\braket q, M, v. = \omega^{q\theta q+x q} =: \psi(q).
	\end{equation*}
	Indeed, choose
	\begin{equation*}
		\theta = 2^{-1} \vartheta, \quad\quad x_i = [v,m_i].
	\end{equation*}
	Using Eq. (\ref{weylCoordinates}), one can then check by direct
	computation that $\psi$ fulfills the defining eigenvalue
	equations
	\begin{equation*}
		\chi([v,m_i]) w(m_i) \psi = \psi
	\end{equation*}
	and hence $\ket\psi.=\ket M,v.$, by Lemma \ref{lmStabilizers}.

	\emph{(\ref{three} $\Rightarrow$ \ref{one}).}
	Reverting the previous proof shows that $\psi$ is a graph state.  It
	has maximal support by definition.
\end{proof}

The claimed analogy between stabilizer states and Gaussian states is
apparent when comparing statement 3 to Theorem \ref{thHudson}.

\section{Mixed States}\label{mixedStates}
\label{scMixedStates}

It is natural to ask how the results obtained before generalize to
mixed states.  Certainly, mixtures of stabilizer states are
non-negative on phase space and it might be surmised that all such
quantum states are convex combinations of stabilizer ones. In the
context of continuous variable systems, Br\"ocker and Werner refuted
an analogous conjecture by giving a counter-example \cite{werner}.
Again, the situation is similar in the finite setting, as will be
shown now.

As a consequence of Theorem
\ref{thWignerProperties}.\ref{wignerParity}, $A(0)$ can be decomposed
as $A(0)=P_+ + P_-$, where $P_\pm$ denotes the projector onto the
symmetric and antisymmetric state vectors respectively.  
Since $P_++P_-=\Id$, we have that $P_- =1/2(\Id - A(0))$. Because we
know the Wigner functions of both $\Id$ ($W(v)=d^{-n}$) and of
$A(0)$ ($W(v)=\delta_{v,0}$), we immediately obtain 
\begin{eqnarray}
	W_{P_-}(v) = \frac12 \left\{
		\begin{array}{ll}
			d^{-n}- 1\quad&v=0 \\
			d^{-n}&\text{else.}
		\end{array}
		\right.
\end{eqnarray}
\begin{figure}
	\label{figFiducial3}
	\centering
	\includegraphics[scale=.5]{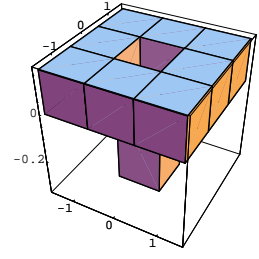}
	\caption{
		Wigner function of the antisymmetric vector $\ket\psi_-.$.
	}
\end{figure} 

\begin{figure}
	\centering
	\includegraphics[scale=.5]{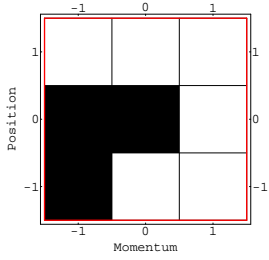}
	\caption{
		\label{figCounter}
		Wigner function of the equal mixture of the vectors $\ket\psi-.,
		w(-1,0)\ket\psi_-.$ and $w(-1,-1)\ket\psi_-.$. White squares stand for
		a value of $1/6$, black squares for $0$.
	}
\end{figure}

For a single three-dimensional quantum system there exists a unique
antisymmetric state vector 
$
	\ket \psi_-.=2^{-1/2}(\ket +1. - \ket -1.),
$
hence $P_- = \ket \psi_-.\bra\psi_-.$.
Figure \ref{figCounter} depicts the Wigner function of the state
$\rho$, obtained by mixing the pure states $\ket\psi_-.,
w(-1,0)\ket\psi_-., w(-1,-1)\ket\psi_-.$ with equal weights.

\begin{figure}
	\begin{tabular}{ccc}
		\includegraphics[scale=.2]{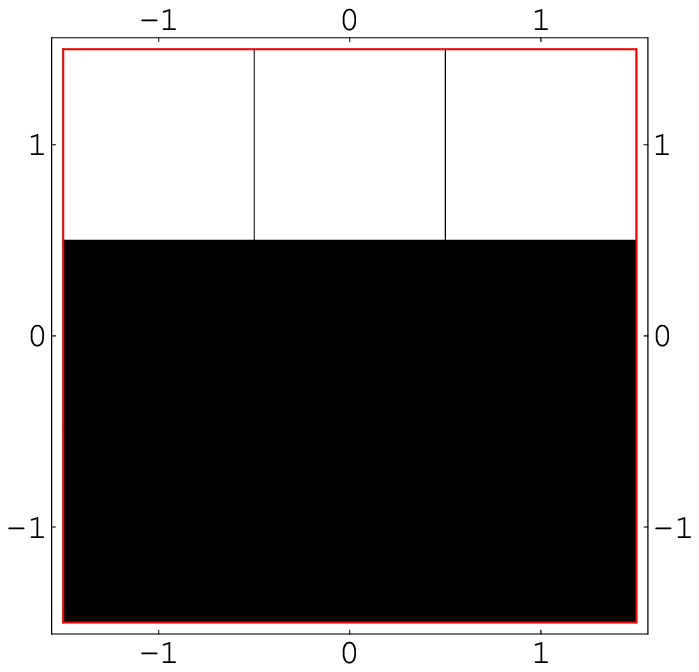}&
		\includegraphics[scale=.2]{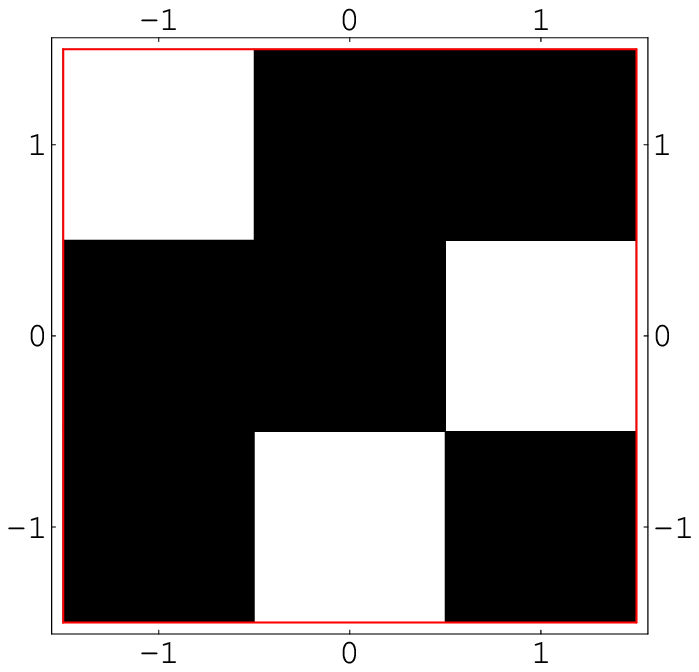}&
		\includegraphics[scale=.2]{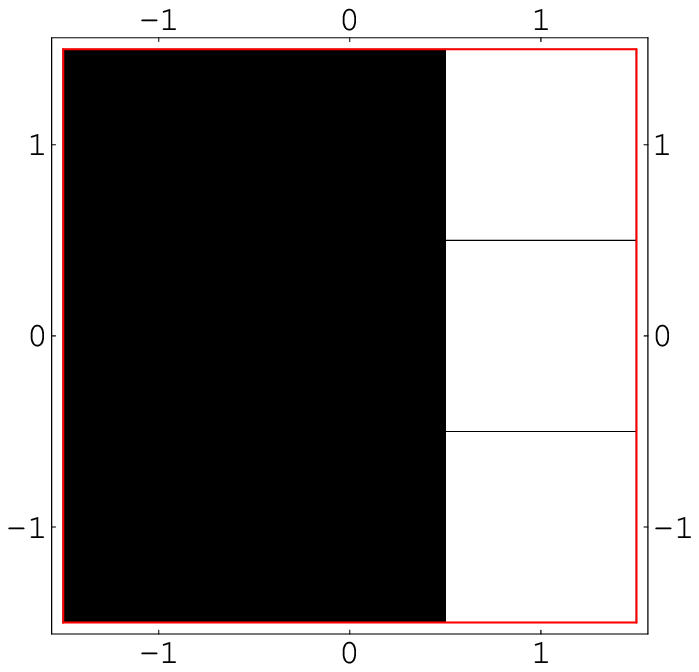}
	\end{tabular}
	\caption{
		\label{figLines}
		The white squares mark all lines in $\ZZ_3^2$ that do not
		intersect any point where the Wigner function shown in Fig.
		\protect{\ref{figCounter}} vanishes.
	}
\end{figure}

The Wigner function of a single-particle stabilizer state is a line in
the two-dimensional phase space, according to Lemma
\ref{lmStabilizerWigner}. There are $d(d+1)$ such lines and hence
equally many stabilizer states. Assume these states have been brought
into some order and denote the associated projection operators by
$P_1, \dots, P_{d(d+1)}$. Let
	$\rho = \sum_i^{d(d+1)} \lambda_i P_i$
be a convex decomposition of $\rho$ in terms these operators.  If
there is a point $v$ in phase space where $W_\rho(v)=0$ and
$W_{P_i}(v)\neq0$, then clearly $\lambda_i$ must vanish. By
exhaustively listing all $12$ lines in $\ZZ_3^2$, one finds that
$\rho$ can have non-zero coefficients only with respect to the
stabilizer states whose Wigner functions are shown in Figure
\ref{figLines}.

But $\rho$ admits no convex decomposition in terms of these three
lines. Indeed, no two of them cover all the points in the support of
$W_\rho$, so only a mixture of all three lines could potentially
suffice. Now notice that the point $(1,-1)$ is an element only of the
third line, while $(1,0)$ is contained in both the second and the
third one. Therefore any mixture of these three lines takes on a
higher value on $(1,0)$ than on $(1,-1)$. The distribution $W_\rho$,
on the other hand, is constant on its support.

\section{dynamics}\label{scDynamics}

Having established which quantum states give rise to non-negative
phase space distributions, the next step is to characterize the set of
operations that preserve this property. We have seen in Section
\ref{scWigner} that Clifford unitaries implement permutations in phase
space and thus manifestly preserve positivity. They are unique in that
regard, as will be shown now.

By the results of Section \ref{scMain}, it is apparent that a
unitary operation $U$ can preserve positivity only if it sends
stabilizer states to stabilizer states.  One can reasonably conjecture
that only Clifford operations possess this feature and in the case of
single-particles in prime-power dimensions, a proof of this fact has
been given in Ref. \cite{galvao}.  The general case, however, poses
surprising difficulties which have forced us to take a less direct
route.

Let us shortly pause to clarify our objectives. We aim to characterize
the set of unitaries $U$ that satisfy statements of the kind:
$W_{U\rho U^\dagger}$ is non-negative whenever $W_\rho$ is. We can
require the above statement to hold for \emph{any} Hermitian operator
$\rho$, or just whenever $\rho$ is a \emph{quantum state}. In the
former case the restrictions on $U$ are much stronger than in the
latter one. Indeed, by considering the image of the phase space point
operators $A(a)$ under the action of $U$ and making use of Lemma
\ref{lmAProps}, it is straight-forward to prove that only Clifford
operations can preserve positivity of the Wigner functions of general
Hermitian operators. The following theorem is slightly more ambitious
in considering only the action of $U$ on quantum states.

\begin{theorem}
	\label{tmDynamics}
	\emph{(Only permutations preserve positivity).}
	Let $U$ be unitary. If, for all quantum states $\rho$ with non-negative
	Wigner function, it holds that $W_{U\rho U^\dagger}$ is non-negative, then
	$U$ is Clifford.
\end{theorem}

\begin{proof}
	Firstly, take a note that substituting 'quantum state' by 'positive
	operator' in the above theorem, only amounts to a change of
	normalization and does not alter the statement. 
	Set
	\begin{eqnarray*}
		\mu(\rho)&:=&\min_{v\in V} W_\rho(v),\\
		\nu(\rho)&:=&\minarg W_\rho := \{v\in V| W_\rho(v)=\mu(v)\}.
	\end{eqnarray*}

	Let $\rho$ be such that $\mu(\rho)<0$. We claim that
	$\mu(\rho)=\mu(\rho')$, where $\rho'=U\rho U^\dagger$. In other
	words: $U$ preserves minimal values.

	Indeed, there exists positive constants $\lambda_{1,2}$ such that 
	\begin{equation*}
		\lambda_1 \mu(\rho') + \lambda_2 d^{-n} = 0.
	\end{equation*}
	Hence $\sigma:=\lambda_1 \rho + \lambda_2 \Id$ has a non-negative Wigner
	function. The assumption $\mu(\rho')<\mu(\rho)$ yields
	\begin{equation*}
		W_{U\sigma U^\dagger}(v) = \lambda_1 \mu(\rho') + \lambda_2 d^{-n} < 0
	\end{equation*}
	for every $v\in \nu(\rho')$, which contradicts the defining property
	of $U$. Thus $\mu(\rho')\leq\mu(\rho)$. Substituting $U$ by $U^{-1}$
	shows that equality of $\mu(\rho)$ and $\mu(\rho')$ must hold.

	Now set 
	\begin{equation*}
		\rho(a):=(1-d^{-n})^{-1}\,w(a) P_-\,w(a)^\dagger
	\end{equation*}
	for all $a\in V$.
	We have $\mu(\rho(a))=\mu(\rho'(a))=-1$ and $\nu(\rho)=\{a\}$. The
	crucial observation lies in the fact that $\nu(\rho')$ contains only
	a single point as well. So, $U$ preserves the 'pointed' shape of
	$W_\rho(a)$. To see why that is the case, suppose there is a $a_0$
	such that $|\nu(\rho(a_0)')|>1$. There are $d^{2n}$ operators
	$\rho(a)'$ and equally many points in phase space, so there exists
	an $a_1$ such that $\nu(a_0)$ and $\nu(a_1)$ intersect in at least
	one point $v$. Define $\sigma=1/2(\rho(a_0)+\rho(a_1))$.  It holds
	that $\mu(\sigma)>-1/2$, whereas $W_{\sigma'}(v)=-1$ which is a
	contradiction.  There is hence a well-defined function $S$ which
	sends $a$ to the unique element of $\nu(\rho(a)')$.  
	
	Finally, let $\sigma$ be any density matrix. The idea is to mix
	$\sigma$ very weakly to $\rho(a)$, so that the positions of the
	minima of the mixture are still determined by $\rho(a)$.
	Indeed, there exists an $\epsilon>0$ such that
	\begin{eqnarray*}
		\nu(\rho(a) + \epsilon \sigma)&=&\{a\}\\
		\mu(\rho(a) + \epsilon \sigma)&=&-1+\epsilon\, W_\sigma(a);\\
		\nu(\rho(a)' + \epsilon \sigma')&=&\{S(a)\}\\
		\mu(\rho(a)' + \epsilon \sigma')&=&-1+\epsilon\, W_\sigma(S(a)).
	\end{eqnarray*}
	Hence $W_{\sigma'}(Sa)=W_\sigma(a)$. We have established that $U$
	acts as a permutation in phase space and is therefore Clifford by
	Lemma \ref{lmAProps}.
\end{proof}

\section{Prime power dimensions}
\label{scPrimePower}

Wigner functions for quantum systems with prime power dimensions have
received particular attention in the literature (most prominently in
Ref. \cite{wootters}). Once again, this is due to the fact that a
finite field of order $d$ exists exactly when $d$ is the power of a
prime and that the field's well-behaved geometrical properties
facilitate many constructions. The present section briefly addresses
the relationship between three natural approaches to Wigner functions
for such systems. We assume the reader is already familiar with the
definition of Weyl operators over Galois fields; a thorough
introduction can be found in Refs. \cite{wootters,diploma}.

Let $d=p^k$ for some prime number $p$. There are three natural ways of
associating a configuration space to $\H$. These are
\begin{enumerate}
	\item\label{primePowerOne}
	an $n$-dimensional vector space over $\ZZ_p$,
	
	\item
	a one-dimensional module over $\ZZ_{p^n}$ or
	
	\item\label{primePowerThree}
	a one-dimensional vector space over the Galois field $\FF_{p^n}$ of
	order $p^n$.
\end{enumerate}
The first and the second of these points of view have manifestly been
covered in this paper. So far we neglected case 3, because -- as we
will see -- it can be completely reduced to the first one.

Let us quickly gather some well-known facts on finite fields.  If $p$
is prime and $n$ a positive integer, $\FF_{p^n}$ denotes the unique
finite field of order $d=p^n$. The simplest case occurs for $n=1$,
when $\FF_p\simeq \ZZ_p$. For $n>1$, fields $\FF_{p^n}$ are realized
by \emph{extending} $\FF_p$, which is then referred to as the
\emph{base field}. Extension fields contain the base field as a
subset.  The extension field possesses the structure of an
$n$-dimensional vector space over the base field. A set of elements of
$\FF_{p^n}$ is a \emph{basis} if it spans the entire field under
addition and $\FF_p$-multiplication. After having chosen a basis
$\{b_1, \dots, b_n\}$, we can specify any element $f=\sum_i f^i b_i$
by its expansion coefficients $\{f^i\}$. The operation 
\begin{equation*}
	\Tr f = \sum_{k=0}^{n-1} f^{p^k}
\end{equation*}
takes on values in the base field and is $\FF_p$-linear. Therefore, 
\begin{equation*}
	\tuple f,g. \mapsto \Tr (fg)
\end{equation*}
defines an $\FF_p$-bilinear form. For any basis $\{b_i\}$,
there exists a \emph{dual basis} $\{b^i\}$ fulfilling the
relation
	$\Tr (b^i b_j) = \delta_{i,j}$
(we do not use Einstein's summation convention). From now on, we
assume that a basis $b_i$ and a dual one $b^i$ have been fixed.

Repeating the construction put forward in Section \ref{scPhaseSpace}, we
introduce the Hilbert space $\H=L^2(\FF_{p^n})$, in other words, $\H$
is the span of $\{\ket q.|q\in \FF_{p^n}\}$. The choice of a basis
induces a tensor structure on $\H$ via
\begin{equation*}
	\ket q.=\ket \sum_i q^i b_i.\mapsto \bigotimes_i \ket q^i..
\end{equation*}

We obtain a character of $\FF_{p^n}$ by setting
$\chi_{p^n}(f)=\chi_p(\Tr f)$. Note that for $n=1$,
$\chi_{p^n}=\chi_p$.  Expanding momentum coordinates $p=\sum_j p_j
b^i$, the character factors:
\begin{equation*}
	\chi(p q) 
	= \chi_p\big(\sum_{i,j} p_j q^i \Tr(b_i b^j)\big)
	= \prod_i \chi_p(p_i q^i).
\end{equation*}
Similarly, the shift and multiply operators factor with respect to
this tensor structure:
\begin{eqnarray*}
	x\big( \sum_i q^i b_i \big) \ket \sum_j x^j b_j.
	&=& \bigotimes_i x^{(i)}(q^i) \ket x^i. \\
	z\big( \sum_i p_i b^i \big) \ket \sum_j x^j b_j. 
	&=& \prod_i \chi_{p}(p_i x^i) \ket \sum_j x^j b_j. \\
	&=& \bigotimes_i z^{(i)}(p_i) \ket x^i., \nonumber
\end{eqnarray*}
where $x^{(i)}$ and $z^{(i)}$ act on the $i$th $p$-dimensional
subsystem. A straight-forward computation along the lines just
presented shows that both the Weyl operators and the phase space point
operators factor:
\begin{eqnarray*}
	w(p,q)
	&=&\bigotimes_i w^{(i)}(p_i,q^i)
	=w(p_1, \dots, p_n, q^1, \dots, q^n) \\
	A(p,q)
	&=&\bigotimes_i A^{(i)}(p_i,q^i)
	=A(p_1, \dots, p_n, q^1, \dots, q^n). \\
\end{eqnarray*}
The above result thus states that the Wigner function induced by the
choice $Q=\FF_{p^n}$ coincides -- up to re-labeling of the phase space
points -- with the one for $Q=\FF_p^n$. In particular, both
definitions give rise to the same set of states with a non-negative
phase space distribution. 

For stabilizer states, however, the situation is not as easy, as will
be discussed subsequently. The preceding discussion suggests defining
a map $\iota: \FF_{p^n}^2 \to \FF_p^{2n}$ by 
\begin{equation*}
	(p,q)\mapsto (p_1, \dots, p_n, q^1, \dots, q^n)
\end{equation*}
(see Refs. \cite{diploma,pittenger}). Let $M$ be a maximal isotropic
subspace of $\FF_{p^n}^2$. It is readily verified that
$\iota(M)\subset \FF_p^{2n}$ is again isotropic and a subspace.
Further, we have shown that the sets of Weyl operators $w(M)$ and
$w(\iota(M))$ coincide and hence so do the stabilizer states $\ket M.$
and $\ket \iota(M).$.

The converse is not true. $\iota^{-1}$ does not necessarily map
$\FF_p^{2n}$ subspaces to those of $\FF_{p^n}^2$. More precisely, if $M\subset
\FF_p^{2n}$ is a subspace, then $\iota^{-1}(M)$ can easily be proven
to be closed under addition, but will in general fail to be closed
under $\FF_{p^n}$-scalar multiplication. This proves the remark made
in the introduction, namely that the set of 'single-particle' (i.e.
$\FF_{p^n}^2$) stabilizer states is a true subset of corresponding 
'multi-particle' set. The following subsection gives a quantitative
account of the relation of the sets.

\subsection{Counting stabilizer codes}

We are going to count the number of stabilizer states of a system
composed of $n$ $d$-level particles. In fact, the computation given
below is slightly more general in that it gives the number of
$k$-dimensional \emph{stabilizer codes} \cite{gottesman}. 

Stabilizer codes are generalizations of stabilizer states. Recall Eq.
(\ref{stabProjector}), where we showed that summing Weyl operators
$w(m)$ over the elements $m$ of a maximal isotropic subspace $M$ of
$V$ yields a one-dimensional projection operator.  It can be shown
that if the requirement of maximality is dropped, the sum still
evaluates to a projector. The range of this operator is the
\emph{stabilizer code} defined by $M$. The dimension $m$ of $M$ and
the dimension $k$ of the stabilizer code are related by $k=d^{n-m}$.

\begin{theorem}
	\emph{(Number of isotropic subspaces)}
	Let $V$ be a $2n$-dimensional symplectic vector space over $\FF_d$,
	where $d$ is the power of a prime. The number of $m$-dimensional
	isotropic subspaces of $V$ is given by
	\begin{equation*}
		\Iso(n,m,d) = \Gauss n,m. \prod_{i=0}^{m-1} (d^{n-i}+1),
	\end{equation*}
	where the square brackets denote the \emph{Gaussian coefficients}
	\begin{equation*}
		\Gauss n,m.=
		\prod_{i=0}^{m-1} \frac{d^{n-i}-1}{d^{m-i}-1}.
	\end{equation*}
\end{theorem}

\begin{proof}
	The proof is inspired by a method employed in Ref.
	\cite{cameron} to solve a related problem. We count the
	number of linearly independent $m$-tuples consisting of mutual
	orthogonal vectors. Indeed, as the first vector $v_1$ we are free to
	choose any non-zero element of $V$. There are $d^{2n}-1$ such
	choices. The second vector must lie in the symplectic complement of
	the span of the first vector	$\tuple v_1.^\bot$.  Hence, $v_2$ can
	be chosen from a $2n-1$-dimensional vector space, the only
	restriction being that $v_2 \not\in \tuple v_1.$. It follows that
	there exist $d^{2n-1}-d^1$ possibilities for $v_2$. Inducting on
	this scheme gives
	\begin{equation}\label{numOfTuples}
		\prod_{i=0}^{m-1} (d^{2n-i}-d^i)
	\end{equation}
	such tuples.

	However, since two different tuples might correspond to the same
	isotropic space, Eq. (\ref{numOfTuples}) over-counted the subspaces. To
	take that fact into account, we must divide by the number of bases
	within an $m$-dimensional space. Arguing in a similar fashion as
	before, we arrive at
		$\prod_{i=0}^{m-1} (d^{m}-d^i)$
	for the sought-for number (see also Ref. \cite{cameron}). Division gives
	\begin{eqnarray*}
		\Iso(n,m,d)
		=
 		\prod_{i=0}^{m-1} 
 			\frac
 				{d^{2n-i}-d^i}
  			{d^{m}-d^i} 
		=
		\prod_{i=0}^{m-1} 
			\frac
				{d^{2(n-i)}-1}
				{d^{m-i}-1}.
	\end{eqnarray*}
	Expanding 
			$d^{2(n-i)}-1 = (d^{n-i}-1)\,(d^{n-i}+1)$
	and using the definition of the Gaussian coefficients concludes the
	proof.
\end{proof}

\begin{corollary}
	The number of $d^{n-m}$-dimensional stabilizer codes defined on $n$
	$d$-level systems is
	\begin{eqnarray*}
		\Stabs(n,m,d)
		&=&
		d^{m} \Gauss n,m.  \prod_{i=0}^{m-1}(d^{n-i}+1).
	\end{eqnarray*}
	In particular, the number of stabilizer states is
	\begin{equation*}
		\Stabs(n,n,d)=d^n \prod_{i=1}^{n}(d^{i}+1).
	\end{equation*}
\end{corollary}

\begin{proof}
	We only need to justify the pre-factor $d^m$.  The defining Eq.
	(\ref{stabProjector}) generates a projector onto a stabilizer code
	given an isotropic space $M$ and a character $\chi([v,\spacedot])$
	on $M$. If $\dim M = m$, then there are $|M|=d^m$ distinct such
	characters (see Appendix \ref{scCharacters}).
\end{proof}

We can now compare the number of stabilizer states for $n$ particles
of dimension $d$ to the corresponding number for a single
$d^n$-dimensional system:
\begin{eqnarray*}
	\frac{\Stabs(n,n,d)}{\Stabs(1,1,d^n)}
	&=& \frac{\prod_{i=1}^n (d^i+1)}{d^n+1} 
	= \prod_{i=1}^{n-1} (d^i+1) \\
	&\geq& d^{\sum_{i=1}^{n-1} i} 
	= d^{\frac12(n^2-n)}. 
\end{eqnarray*}
This is the super-exponential scaling mentioned in the introduction.


\section{Acknowledgments}

The author is grateful for support and advice provided by Jens Eisert
during all stages of this project.

Thanks to Dirk Schlingemann for enlightening conversations on phase
space techniques.  The figures were produced using \emph{Mathematica}
notebooks \cite{diploma} partly based on Timo Felbinger's {\tt
qmatrix} package \cite{timo}. Martin Plenio and Alessio Serafini gave
helpful comments on draft versions of this paper. Useful references
were pointed out to the author by S. Chaturvedi, C. K. Zachos,  A. Klimov, 
and M. Ruzzi.

This work has benefitted from funding provided by the European
Research Councils (EURYI grant of J. Eisert), the European Commission
(Integrated Project QAP), the EPSRC (Interdisciplinary Research
Collaboration IRC-QIP), and the DFG.

\section{Appendix}

\subsection{Discrete Stone-von Neumann Theorem}
\label{scCliffordProof}

This section generalizes well-known results for prime-power dimensions
(see e.g. Ref. \cite{neuhauser} and citations therein) to all odd $d$.
The proof is based on some simple observations employing group
representation theory. We state a preparing lemma beforehand.

\begin{lemma} 
	The Weyl representation is irreducible.
\end{lemma}

\begin{proof}
	We compute
	\begin{eqnarray*}
		\frac1{|H(\ZZ_d^n)|} \sum_{{a\in V,}\atop{t\in \ZZ_d}} |\tr w(a,t)|^2 
		&=& d^{-(2n+1)} \sum_t |\tr w(0,t)|^2 \\
		&=& d^{-(2n+1)} \sum_t d^{2 n} = 1
	\end{eqnarray*}
	which establishes irreducibility by a well-known criterion from
	group representation theory (see any textbook on that topic, e.g.
	\cite{elementarygrouptheory}).
\end{proof}

\begin{proof} \emph{(of Theorem \ref{thCliffordStructure})}
	By the composition law Eq. (\ref{heisenbergComp}) it is clear
	that $w'(p,q,t):=w(S(p,q),t)$ is a representation of the Heisenberg
	group which affords the same character (i.e. $\tr w(a,t)=\tr w'(a,t)$).
	The preceding lemma
	yields that $w$ and $w'$ are equivalent and thus the existence of
	$\mu(S)$ follows. Further,
	\begin{eqnarray*}
		\mu(S)\mu(T) w(p,q) \mu(T)^\dagger\mu(S)^\dagger
		&=& \mu(S) w(T(p,q)) \mu(S)^\dagger \\
		&=& w(S\,T(p,q)) \\
		&=& \mu(ST) w(p,q) \mu(ST)^\dagger.
	\end{eqnarray*}
	Because the Weyl matrices span the set of all operators, the last
	line fixes $\mu(ST)$ modulo a phase and we have proven the second
	assertion.
	
	We turn to the last claim. Let $S$ and $c$ be as defined in Eq.
	(\ref{cliffordDefinition}).  Using the commutation relations Eq.
	(\ref{heisenbergComp}) and the fact that conjugation by unitaries
	leaves the center $\chi(t) \Id$ of the Weyl representation
	invariant, it is easy to see that $S$ must be an isometry in the sense
	that $[Sa,Sb]=[a,b]$. 
	To proceed, consider the following calculation. On the one hand
	\begin{eqnarray}
		U w(a) w(b) U^\dagger 
		&=& U w(a+b,2^{-1}[a,b]) U^\dagger \label{ii} \\
		&=& w(S(a+b),2^{-1}[a,b]) c(a+b), \nonumber 
	\end{eqnarray}
	while on the other hand,
	\begin{eqnarray}
		U w(a) w(b) U^\dagger 
		&=& U w(a) U^\dagger U w(b)\label{v}\\
		&=& w(Sa) w(Sb) c(a) c(b) \nonumber \\
		&=& w(Sa+Sb,2^{-1}[Sa,Sb]) c(a) c(b). \nonumber  
	\end{eqnarray}
	Comparing the last lines of Eqs. (\ref{ii}) and (\ref{v}) one finds
	that  $S$ must be compatible with addition in $\ZZ_d^{2n}$ meaning
	that $S(a+b)=Sa+Sb$.  Because $\ZZ_d$ is \emph{cyclic} the preceding
	property implies that $S$ is also compatible with scalar
	multiplication: 
	\begin{eqnarray*}
		S(\lambda a)=S(a+\cdots+a)=S(a)+\cdots+S(a)=\lambda S(a).
	\end{eqnarray*} 
	Hence $S$ is linear and therefore symplectic.  Lastly, again using
	lines (\ref{ii}) and (\ref{v}), we have that $c(a+b)=c(a)c(b)$ and
	conclude that $c$  is a character. 
	By Lemma \ref{lmCharacters}, there exists an $a_0\in V$ such that
	$c(\spacedot)=\chi([a_0,S\spacedot])$.
	Thus:
	\begin{eqnarray*}
		w(a_0) \mu(S)\,w(a)\,\mu(S)^\dagger w(a_0)^\dagger 
		&=& w(a_0) w(Sa) w(-a_0) \\
		&=& \chi([a_0, Sa]) w(Sa) \\
		&=& c(a) w(Sa).
	\end{eqnarray*}
\end{proof}

\subsection{Axiomatic Characterization of the Wigner function}
\label{scUniqueness}

The discussion in Section \ref{scWigner} should suggest that
Definition \ref{dfWigner} yields 'the' natural analogue of the
original continuous Wigner function. However, to bolster that claim
with more objective arguments, we establish that -- at least in prime
dimensions -- the form is virtually determined by the property of
Clifford covariance (Theorem \ref{thWignerCovariance}). 

\begin{theorem}
	\emph{(Uniqueness)}
	Let $d$ be an odd prime. Let $Q,V,\H$ be as usual.  Consider a mapping $W'$
	that fulfills the following axioms.
	\begin{enumerate}
		\item \emph{(Phase space)}
		$W'$ is a linear mapping sending operators to functions on
		the phase space $V$.

		\item
		\emph{(Clifford covariance)}
		$W'$ is covariant under the action of the Clifford group, in the
		sense of Theorem \ref{thWignerCovariance}.
	\end{enumerate}
	Then
		$W_\rho'(p,q) = \lambda_1 W_\rho(p,q) + \lambda_2$
	for two constants $\lambda_{1,2}$.  If further, 
	\begin{enumerate}
		\setcounter{enumi}{2}
		\item \emph{(Marginal probabilities)}
		$W'$ gives the correct marginal probabilities, as stated in 
		Theorem \ref{thWignerProperties}.\ref{wignerMarginal},
	\end{enumerate}
	then $W'(p,q)=W(p,q)$.
\end{theorem}

\begin{proof}
	Consider an alternative definition $\rho \mapsto W'_\rho$ of a
	Wigner function. Linearity implies the existence of a set of
	operators $\{A'(v)\}$ such that
	$
		W'(v)=d^{-n} \tr(A'(v) \rho)
	$.
	$W'$ is covariant under the action of the Weyl operators if and only
	if $A'(v)=w(v)A'(0)w(v)^\dagger$. So the only degree of freedom left
	in the definition of $W'$ is the choice of $A'(0)$. Again, one must
	require $A'(S\,v)=\mu(S)\,A(v)\,\mu(S)$ if Theorem
	\ref{thWignerCovariance} is to hold. In
	particular, because the origin $0$ is a fixed point of any linear
	operation, $A'(0)$ must commute with all $\mu(S)$.

	As a consequence, the old, unprimed Wigner function $W_{A'(0)}$ of
	$A'(0)$ stays fixed under any symplectic operation $S$.	Since any two
	non-zero points of $V$ can be mapped onto each other by a suitable
	symplectic matrix $S$, $W_{A'(v)}$ must be constant on all such
	points.  So there are only two parameters free to be chosen:
	$W_{A'(0)}(0)$ and $W_{A'(0)}(v), v\neq0$.  Clearly, the set of all
	operators that comply with these constraints is spanned by $\Id$ and
	$A(0)$: 
	\begin{equation}\label{uniqueForm}
		A'(0)=\lambda_1\,\Id+\lambda_2\,A(0).
	\end{equation}
	The above decomposition implies the first statement of the Theorem.

	As for the second claim, choose an $a\in V$. The projection operator
	$\ket a.\bra a.$ is invariant under the action of Weyl operators of
	the form $w(p,0)$. Thus, due to Clifford covariance, the Wigner
	function $W'_{\ket a.}$ must be $p$-shift invariant: $W'_{\ket
	a.}(p+p',q)=W'_{\ket a.}(p,q)$. We required Theorem
	\ref{thWignerProperties}.\ref{wignerMarginal} to hold, hence
	\begin{equation*}
		\sum_{p\in Q} W'_{\ket a.}(p,0) = d^n\,W'(0,0) = \delta_{a,0}.
	\end{equation*}
	By Eq. (\ref{uniqueForm}) and Theorem
	\ref{thWignerProperties}.\ref{wignerParity} it follows that
		$W'(0,0) 
		= d^{-n}(\lambda_1 + \lambda_2 \delta_{a,0})$,
	yielding $\lambda_1=0, \lambda_2=1$.
\end{proof}


\subsection{Characters and Complements}
\label{scCharacters}

Consider a space $R=\ZZ_d^n$ with a bilinear form $\tuple \spacedot,
\spacedot.: R\times R\to \ZZ_d$. For any $s\in R$ the function
$r\mapsto \chi(\tuple s, r.)$ defines a character of $R$. The form is said
to be \emph{non-degenerate} if $\tuple s, \spacedot. \neq \tuple
s',\spacedot.$ for distinct $s,s'$. The two spaces we are concerned with
are $Q$ with the canonical scalar product and $V$ with the symplectic
scalar product. Both can easily be checked to be non-degenerate.

The following lemma states a basic fact about spaces with non-degenerate
forms. We repeat it for completeness.

\begin{lemma}\label{lmCharacters}
	Let $R=\ZZ_d^n$ with non-degenerate bilinear form
	$\tuple\spacedot,\spacedot.$.
	Any character $\zeta$ of $R$ is of the form $\zeta(r)=\chi([s,r])$
	for some unique $s\in R$.
\end{lemma}

\begin{proof}
	Addition gives $V$ the structure of a finite abelian group.
	Therefore, $V\simeq V^*$, as is well-known (see e.g. Ref.
	\cite{harmonictextbook}). So there are $|V|$ different
	characters of $V$, but equally many of the form $\chi([v,\spacedot])$.
\end{proof}

If $d$ is prime and $M$ a subspace of $V$, the well-known relation
$\dim M+\dim M^\bot = \dim V$ holds \cite{huppert}. It is, however, no
longer true in the general case. A counter-example can be constructed
along the same lines as in Section \ref{scFourier}. Still, an analogue
exists as demonstrated below.

\begin{theorem}\label{thComplements}
	Let $R=\ZZ_d^n$ with non-degenerate bilinear form
	$\tuple\spacedot,\spacedot.$. If $M$ denotes a subspace of $R$, then the 
	'complementarity relation' $|M|\,|M^\bot|=|R|$ holds. 
\end{theorem}

\begin{proof}
	We will show that
	\begin{equation}\label{complementDuality}
		M^\bot \simeq \big(V/M\big)^*.
	\end{equation}

	For $m\in M^\bot$, the relation
		$[v] \mapsto \chi([m,v])$
	defines a character of $V/M$, as can easily be verified. Let us
	denote the map $m\mapsto \chi([m,\spacedot])$ by $\iota_1$.

	Conversely, given an element $\zeta$ of $\big(V/M\big)^*$,
		$v \mapsto \zeta([v])$
	is a character of $V$. By Lemma \ref{lmCharacters} there exists a
	unique $w\in V$ such that $\zeta([v])=\chi([w,v])$. If $m\in M$,
	then $\zeta([m])=\zeta([0])=1$ and hence $w\in M^\bot$. Using the
	notions just introduced, we can define $\iota_2: \big(V/M\big)^* \to
	M^\bot$ by $\zeta \mapsto w$.

	It is simple to check that $\iota_2=\iota_1^{-1}$. In particular,
	$\iota_1$ is invertible and Eq. (\ref{complementDuality}) follows.

	With the help of Lagrange's Theorem, we can compute
	\begin{eqnarray*}
		\left| M^\bot \right| 
		= \left| \big(V/M\big)^* \right| 
		= \left| V/M \right| 
		= |V|/|M|,
	\end{eqnarray*}
	which concludes the proof.
\end{proof}

\begin{corollary}
	\label{crComplements}
	Let $V, Q$ be defined as usual. Let $M$ be an isotropic subspace of $V$ and
	$S$ be any subspace of $Q$.

	\begin{enumerate}
		\item\label{symplecticComplement}
		\emph{(Maximally isotropic spaces)}
		$M$ is equal to its symplectic complement $M^\bot$ if and only if
		$|M|=d^n$.

		\item\label{orthogonalComplement}
		\emph{(Characters of subspaces)}
		Any character $\zeta$ of $S$ can be written as $\zeta(s)=\chi(qs)$
		for a suitable $q\in Q$.
	\end{enumerate}
\end{corollary}

\begin{proof}
	Claim \ref{symplecticComplement} follows immediately from Theorem
	\ref{thComplements} and the fact that isotropic spaces are contained
	in their symplectic complement: $M\subset M^\bot$.

	We turn to the second statement. In Lemma \ref{lmWignerConstant} we
	have argued that the characters of $S$ which are expressable as
	$\chi(qs)$ stand in one-to-one correspondence to cosets in
	$Q/S^\bot$. But $|Q/S^\bot|=|S|$ and hence all characters are of
	that form.  \end{proof}

\subsection{A geometric note}
\label{scGeometric}

The proof of the Main Theorem makes use of the fact that for any
vector $v \in V$, there exists a symplectic operation $S$ that sends
$v$ to a vector of the form $(p,0)$.  Indeed, if $d$ is prime, any two
vectors are similar, in the sense that they can be mapped onto each
other by a symplectic matrix. Technically, this is a trivial
incarnation of Witt's Lemma (see Ref.  \cite{aschbacher} for a
formulation that is applicable in our context). 

Once again the non-prime case poses additional difficulties.  Recall
that the \emph{order} of a $v\in V$ is the least positive $\lambda \in
\ZZ_d$ such that $\lambda\,v=0$. It is easy to see that the order of a
vector is left invariant by the action of invertible linear mappings.
If $d$ is a composite number (i.e. not prime), then $V=\ZZ_d^{2n}$
contains elements of different orders which cannot be related by a
linear operation. However, one might conjecture that any two vectors of
equal order are similar. This is the content of the following lemma.
Some concepts used in the proof can be found in Refs.
\cite{huppert,zmud}.

\begin{lemma} \emph{(Similarity)}
	Let $V=\ZZ_d^{2n}$. Let $a_1, a_2 \in V$ be two vectors with the same
	order. Then there exists a symplectic matrix $S$ such that
	$S a_1=a_2$.
\end{lemma}

\begin{proof}
	We can slightly weaken the assumptions made about $V$. All we
	require for this proof is that $V$ is a finite $\ZZ_d$-module 
	with non-degenerate symplectic form $[\spacedot,\spacedot]$. It need
	not be of the form $\ZZ_d^{2n}$. 
	
	Let $v\in V$ be a vector of order $d$. As $v\mapsto
	\chi([v,\spacedot])$ implements an isomorphism, $V\to V^*$,
	$\ord\big(\chi([v,\spacedot])\big)=\ord(v)=d$. There hence exists a
	$w\in V$ such that $[v,w]=\lambda$ has order $d$. Any such number
	possesses a multiplicative inverse $\lambda^{-1}$ modulo $\ZZ_d$ and
	hence $w'=\lambda^{-1}$ fulfills $[v,w']=1$. Vectors satisfying such
	a relation are said to be \emph{hyperbolic couples}.  Denote their
	span $\tuple \{v,w'\}.$ as $H$.

	Set $V':=H^\bot$. By Theorem \ref{thComplements} $|V|=|H|\,|V'|$.
	Further, it is easy to see that $H^\bot\cap H=\{0\}$ and hence
	$V=H\operp V'$, where $\operp$ denotes the \emph{orthogonal direct
	sum}.  We claim that the symplectic inner product is non-degenerate
	on $V'$. Indeed, suppose there is a non-zero $v'\in V'$ such that
	$[v',w']=0$ for all $w'\in V'$. Then, by definition of $V'$,
	$[h,w']=0$ for all $h\in H$ and therefore $v'$ would be orthogonal 
	on all vectors of $V$. Hence such a $v'$ cannot exist by the
	non-degeneracy of $[\spacedot,\spacedot]$.

	Note that $V'$ fulfills the assumptions made about $V$ at the beginning
	of the proof and has strictly smaller cardinality. Thus, we can
	induct on $|V|$ to obtain a decomposition
	\begin{equation*}
		V=H_1\operp\dots\operp H_n
	\end{equation*}
	of $V$ in terms of two-dimensional subspaces spanned by hyperbolic
	couples $\{v_i, w_i'\}$.
	We arrange these vectors as the columns of a matrix
	$S=(v_1,\dots,v_n,w_1',\dots,w_n')$. The construction of the couples
	$\{v_i, w_i'\}$ ensures that $S$ is symplectic, as can easily be
	verified. 

	Now let $a_1, a_2\in V$ be two vectors with maximal order. By the
	preceding discussion, there exists symplectic matrices $S_i$ having
	$a_i$ as their respective first column. Clearly, then $S_2 S_1^{-1}
	a_1 = a_2$.

	Lastly, suppose $\ord(a_i)=k\leq d$. It is easy to see that
	$a_i'={k a_i}/d$ are elements of $V$ with maximal order. Further, if
	$S$ maps $a_1'$ to $a_2'$, then also $a_1$ to $a_2$.
\end{proof}

\begin{corollary}\label{corTransitive}
	\emph{(Transitive action)}
	Let $\ket M_1, v_1.$, $\ket M_2, v_2.$ be stabilizer states.
	If their respective associated isotropic subspaces $M_1, M_2$ are
	spanned by vectors of maximal order, then there exists a Clifford
	operation relating these state vectors.
\end{corollary}

\begin{proof}
	Let $\{m^{(i)}_1, \dots, m^{(i)}_n\}$, $i = 1,2$ be bases of $M_1$
	and $M_2$ respectively. Assume that all vectors have maximal order.
	It is simple to adapt the previous proof for constructing a
	symplectic matrix $S$ sending $m^{(1)}_i$ to $m^{(2)}_i$. 
\end{proof}


\subsection{Some properties of the phase space point operators}
\label{scMoyal}

\begin{lemma}
	\label{lmAProps}
	\emph{(Properties of the phase space point operators)}
	The phase space point operators fulfill the following relations
	\begin{eqnarray*}
		A(a)&=&w(2a)A(0),\\
		A(a)A(b)&=&w(2a-2b),\\
		\tr(A(u)\,A(v)\,A(w)) &=& \chi([v,u]+[u,w]+[w,v]).
	\end{eqnarray*}

	Further, if $U$ permutes the phase space point operators under
	conjugation
	\begin{equation*}
		U A(v) U^\dagger = A(v')
	\end{equation*}
	for all $v\in V,$ 
	then $U$ is Clifford.
\end{lemma}

\begin{proof}
	Clifford covariance (Theorem \ref{thWignerCovariance}) implies
	$A(a)=w(a)A(0)w(a)\dagger$. 
	Using Theorem \ref{thWignerProperties}.\ref{wignerParity} it is 
	easy to see that $A(0)w(a)A(0)=w(-a)$ and $A(0)^2=\Id$. 
	Hence
	\begin{equation*}
		A(a)=w(a)A(0)w(-a)A(0)\,A(0)=w(2a)A(0)
	\end{equation*}
	proving the first relation. The second one follows.
		
	For the proof of the third equation, we abbreviate 
	$A(0)$ as $A$. Then
	\begin{eqnarray*}
		&&\tr\left(A(u) A(v) A(w)\right) \\
		&=&\tr(w(2u) A\, w(2v) A\, w(2w) A) \\
		&=&\tr(w(2u) w(-2v) w(2w) A^3) \\
		&=&\chi([u,-v]+[u-v,w]) \tr(w(2(u-v+w)) A) \\
		&=&\chi([v,u]+[u,w]+[w,v]) \tr(A(u-v+w)).
	\end{eqnarray*}
	It has been noted in Theorem
	\ref{thWignerProperties}.\ref{wignerOrthonormal} that phase space
	point operators have unit trace, which concludes the proof.

	Lastly, suppose the action of $U$ permutes phase space point
	operators. For any $a\in V$, we have
	\begin{eqnarray*}
		U w(a) U^\dagger 
		&=& U w(2 \, 2^{-1}(a-0)) U^\dagger \\
		&=& U A(a) U \, U^\dagger A(0) U^\dagger \\
		&=& A(a') A(0') \\
		&=& w(2(a'-0'))
	\end{eqnarray*}
	for suitable $a',0'\in V$. Hence $U$ maps Weyl operators to Weyl
	operators and is thus Clifford by definition.
\end{proof}

\end{document}